\documentclass{article}

\usepackage{arxiv}

\usepackage[utf8]{inputenc} 
\usepackage[T1]{fontenc}    
\usepackage{hyperref}       
\usepackage{url}            
\usepackage{booktabs}       
\usepackage{amsfonts}       
\usepackage{nicefrac}       
\usepackage{microtype}      
\usepackage{lipsum}
\usepackage{graphicx}
\graphicspath{ {./images/} }

\title{Classifying daily activities needs posture, reconstructing them needs motion}

\author{
 Arefeh Farahmandi \\
 Centre for Neuroscience Studies\\
 Queen's University\\
 Kingson, Ontario, Canada\\
  \texttt{21afna@queensu.ca} \\
   \And
 Gunnar Blohm \\
Centre for Neuroscience Studies\\
 Queen's University\\
 Kingson, Ontario, Canada\\
\texttt{gunnar.blohm@queensu.ca} \\
  \And
}

\begin{document}
\maketitle
\begin{abstract}

Humans recognize movements effortlessly, even from noisy and complex visual input. But what information in the stimulus allows humans to rapidly classify movements from the high-dimensional space of body configurations? No computational framework has systematically compared different strategies of movement analysis to address this question. Here, we used videos of 16 daily activities from the MoVi dataset~\cite{Ghorbani2020MoVi:Dataset}, spanning both similar and highly distinct motion types, and compared three strategies: Temporal Movement Primitives (TMPs), which decompose movements into weighted sums of temporally smooth basis functions; Legendre polynomial coefficients, which project joint-coordinate trajectories onto an orthogonal polynomial basis; and autoencoder latent embeddings. Legendre coefficients and TMPs achieved the highest random forest classifier accuracy (96\%), followed by autoencoders (89\%). We found two discriminative features for movement classification. The first and most informative is the general posture of the body, the average spatial configuration that distinguishes one activity from another. Additionally, we identified 9 (out of 16 possible) joints (Wrists , Elbows, Knees, Neck, Ankles) that are most predictive for movement classification. Interestingly, good classification accuracy did not automatically lead to good movement generation: when we reconstructed movements for each activity, TMPs preserved the temporal dynamics and produced perceptually natural motion, whereas reconstructions from Legendre coefficients retained only the average posture and appeared frozen and unnatural. Together, these results reveal a dissociation in how movement information is organized: the static configuration of the body suffices to classify what activity is performed, but the temporal dynamics of movement are required to reconstruct how it unfolds. This distinction clarifies which features the visual system may rely upon for rapid action recognition, and suggests that compact, interpretable postural features could enable efficient movement screening in clinical and monitoring applications, while dynamic representations remain essential wherever faithful movement generation is the goal.
\end{abstract}


\section{Introduction}
Despite the complexity and variability of sensory input, the human visual system perceives and categorizes different movements with remarkable speed and accuracy. Point-light display
experiments established decades ago that observers perceive biological motion from minimal kinematic
information and suffice for action recognition, gender
discrimination, and even emotion attribution ~\cite{Johansson1973VisualAnalysis,Troje2002DecomposingPatterns,Pollick2001PerceivingMovement}. This
perceptual efficiency implies that the visual system extracts compact, informative features from the high-
dimensional space of possible body configurations and their temporal evolution. A central question for both
computational neuroscience and clinical movement analysis is: what are these features, and can we identify
them computationally?

The challenge of characterizing movement mathematically is rooted in what Bernstein~\cite{Bernstein67CoordinateMovements} identified as the
degrees-of-freedom problem: the human body possesses far more kinematic degrees of freedom than any single
task requires. One influential solution is the motor modularity hypothesis, which proposes that the central
nervous system constructs complex movements from a limited repertoire of building blocks known as motor primitives
or muscle synergies~\cite{Bizzi1991ComputationsPerspective,Flash2005MotorInvertebrates,Giszter2015MotorQuestions}. Foundational work
demonstrated that spinal microstimulation in frogs evokes endpoint forces that combine linearly to
produce limb movements~\cite{Mussa-Ivaldi1994LinearControl},
and a small number of muscle synergies account for the diversity of both amphibian and human limb
behaviors~\cite{Tresch1999TheCord,DAvella2003CombinationsBehavior,DAvella2006ControlCombinations}. These findings, consolidated across species and
organizational levels~\cite{Bizzi2008CombiningMovement,Giszter2015MotorQuestions}, established the framework of modular motor control.
Yet different decomposition methods (spatial, temporal, spatiotemporal) capture different aspects of
movement structure ~\cite{Chiovetto2022TowardPrimitives}. So, the question of which strategy best serves classification,
reconstruction and interpretation of movement remains open.

Several analysis strategies have been proposed, each embodying different assumptions about movement
structure. Temporal Movement Primitives (TMPs) model movements as weighted sums of temporally smooth
basis functions learned under Gaussian process priors, with the number of primitives determined by Bayesian model selection via a Laplace approximation to the model evidence~\cite{Endres2013ModelPrimitives}. This principled
approach avoids heuristic parameter choices and produces perceptually
valid movement reconstructions ~\cite{Knopp2019PredictingModels,Leh2023AValidity}. In another view, functional data
analysis represents trajectories as projections onto basis functions, with coefficients serving as features~\cite{Ramsay2005FunctionalAnalysis,Warmenhoven2021PCABiomechanics}. Within this family, Legendre polynomials offer a uniquely
interpretable basis: each polynomial degree maps onto a distinct biomechanical meaning: degree zero
captures the static posture, while degrees one and two capture the simplest
temporal dynamics. While Legendre bases have been used for continuous-time memory in neural networks~\cite{Voelker2019LegendreNetworks} and polynomial basis functions appear in robotics trajectory parameterizations~\cite{Paraschos2013ProbabilisticPrimitives}, their application to human movement classification has not been explored. A third approach leverages
data-driven deep learning: autoencoders learn compressed latent features through a reconstruction
objective without explicit structural assumptions~\cite{Holden2016AEditing,Butepage2017DeepClassification}. However,
features extracted for reconstruction are not necessarily optimal for discrimination as handcrafted
features have been shown to generalize better than deep features for activity recognition~\cite{Bento2022ComparingRecognition}. Also, standard variational autoencoders conflate task-relevant and task-irrelevant variation without explicit
disentanglement~\cite{Noseworthy2020Task-ConditionedPrimitives}. This tension between reconstructive fidelity and discriminative
power is fundamental: a strategy that faithfully reproduces all aspects of movement may dilute the
specific features that distinguish one action from another. More broadly, while state-of-the-art deep learning methods such as graph convolutional networks achieve impressive classification accuracy on action recognition benchmarks ~\cite{Yan2019AData}, they function as black boxes that do not reveal which movement features drive discrimination, limiting their value for scientific insight into movement organization.

This tension connects to a long-standing debate in biological motion perception: the relative
importance of form (static body configuration), versus motion (temporal dynamics) for action
recognition. While Johansson's ~\cite{Johansson1973VisualAnalysis} original demonstrations emphasized motion cues, subsequent work
revealed that form alone carries substantial discriminative information. Beintema and Lappe ~\cite{Beintema2002PerceptionMotion} showed that
observers recognize walking even when local motion signals are eliminated, and Lange and Lappe ~\cite{Lange2006ACues}
proposed a model where recognition proceeds through posture template matching without requiring temporal
order; spatial scrambling devastated performance while temporal scrambling had no effect. Giese and Poggio
~\cite{Giese2003NeuralMovements} formalized this in a dual-pathway architecture, with ventral form and dorsal motion pathways supported
by neuroimaging evidence for a double dissociation~\cite{Vangeneugden2014DistinctDiscriminations}. Converging evidence from
computer vision confirms that single-frame spatial information achieves substantial action recognition accuracy~\cite{Simonyan2014Two-StreamVideos}. Critically, the relative importance of form versus motion depends on the level
of discrimination: form dominates when distinguishing between fundamentally different actions, while
dynamics become essential for within-action discriminations such as gender or emotion~\cite{Troje2002DecomposingPatterns,Roether2009CriticalGait}. Understanding which movement features enable classification, and whether computational
methods recover the same features that human observers rely upon, is not purely academic. As movement
analysis enters clinical practice through markerless pose estimation systems~\cite{Stenum2024ClinicalChange,Uhlrich2023OpenCap:Videos}, the demand for interpretable, biomechanically meaningful features has become urgent~\cite{Rudin2019StopInstead,Xiang2025ExplainableReview,Slijepcevic2023ExplainablePalsy}. Yet despite the independent maturation of motor primitive theory,
polynomial trajectory analysis, deep learning, and biological motion perception research, no study has
systematically compared these strategies to determine which movement features drive activity classification and how classification relates to reconstruction fidelity, and whether the features that enable
computational discrimination align with those that support human movement perception.

In the present study, we compared three movement analysis strategies (Temporal
Movement Primitives, Legendre polynomial coefficients, and autoencoder latent embeddings) on 16 daily
activities from the MoVi dataset~\cite{Ghorbani2020MoVi:Dataset}. For each strategy, we evaluate cross-validated
classification accuracy, reconstruction quality of generated movements, and interpretability of the learned
features. We systematically vary the complexity of each method to identify optimal parameters and investigate the impact of extracted features on classification performance and reconstruction quality. Critically, we decompose the contribution of postural versus dynamic information and examine how
classification changes. Also, we identify anatomical joints which carry the
most discriminative information. Our approach is driven by two complementary goals: to provide practical guidance on which strategy best serves different applications, and to characterize how discriminative movement information is organized, in particular the relative contributions of postural configuration and temporal dynamics. We then discuss how this computational organization relates to the form-versus-motion distinction described in biological motion perception.

\section{Materials and methods}

\begin{figure}[!htbp]
\centering
\includegraphics[width=\textwidth]{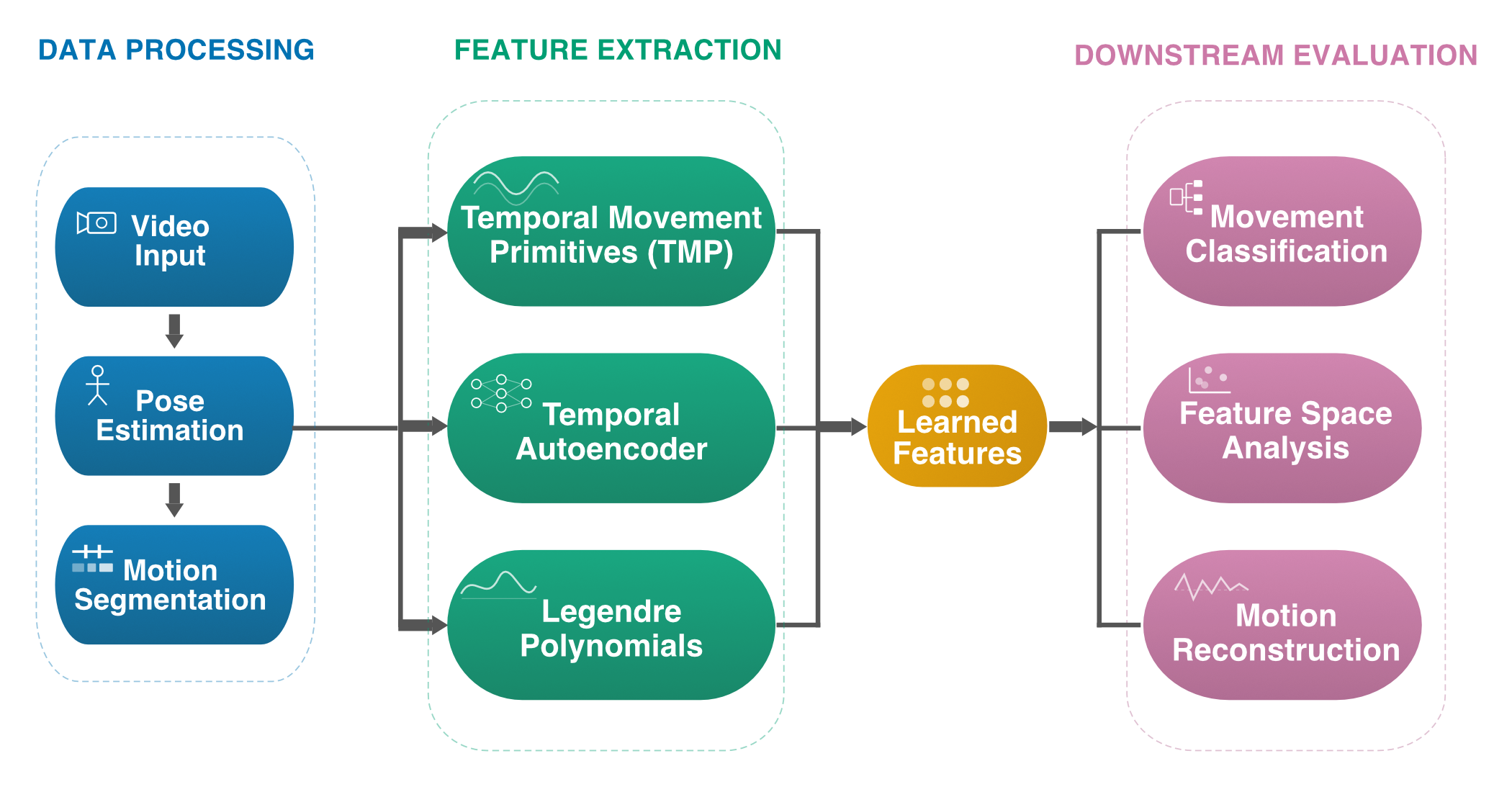}
\caption{\textbf{Overview of the movement analysis framework.}
The pipeline consists of three stages. Data processing converts raw video 
recordings into segmented joint-position trajectories via pose estimation 
(MMPose, 16 joints) and manual motion segmentation. Feature extraction 
fits three parallel models to all segments and produces a fixed-length feature vector set for each method. Downstream evaluation 
assesses the resulting features through movement classification across 
16 daily activities, feature space analysis (dimensionality reduction, 
posture-dynamics decomposition, joint selection), and motion 
reconstruction from both category-averaged and per-segment features.}
\label{fig1}
\end{figure}

In this work, we investigated three different strategies (Temporal
Movement Primitives, Legendre polynomial coefficients, and autoencoder latent embeddings) to understand what features enable humans to perceive and classify different movements (Fig~\ref{fig1}). To achieve this goal, we carried out a series of steps, each building on the previous one. First, since our strategies require joint-level input rather than raw video, we describe how we processed the videos to extract 3D joint positions and segment continuous recordings into individual movement cycles (Data Processing). Next, to compare which strategy produces the most discriminative features, we fit three parallel models to the segmented joint-position trajectories: Temporal Movement Primitives (TMPs), Legendre polynomial fitting, and an autoencoder; each embodies different assumptions about movement structure and is detailed in the Feature Extraction section. We then compared the three strategies and evaluated what they reveal about movement through a series of downstream analyses (Downstream Evaluation). Below, we detail each of these steps.

\subsection{Data processing}

\begin{figure}[!htbp]
\centering
\includegraphics[width=\textwidth]{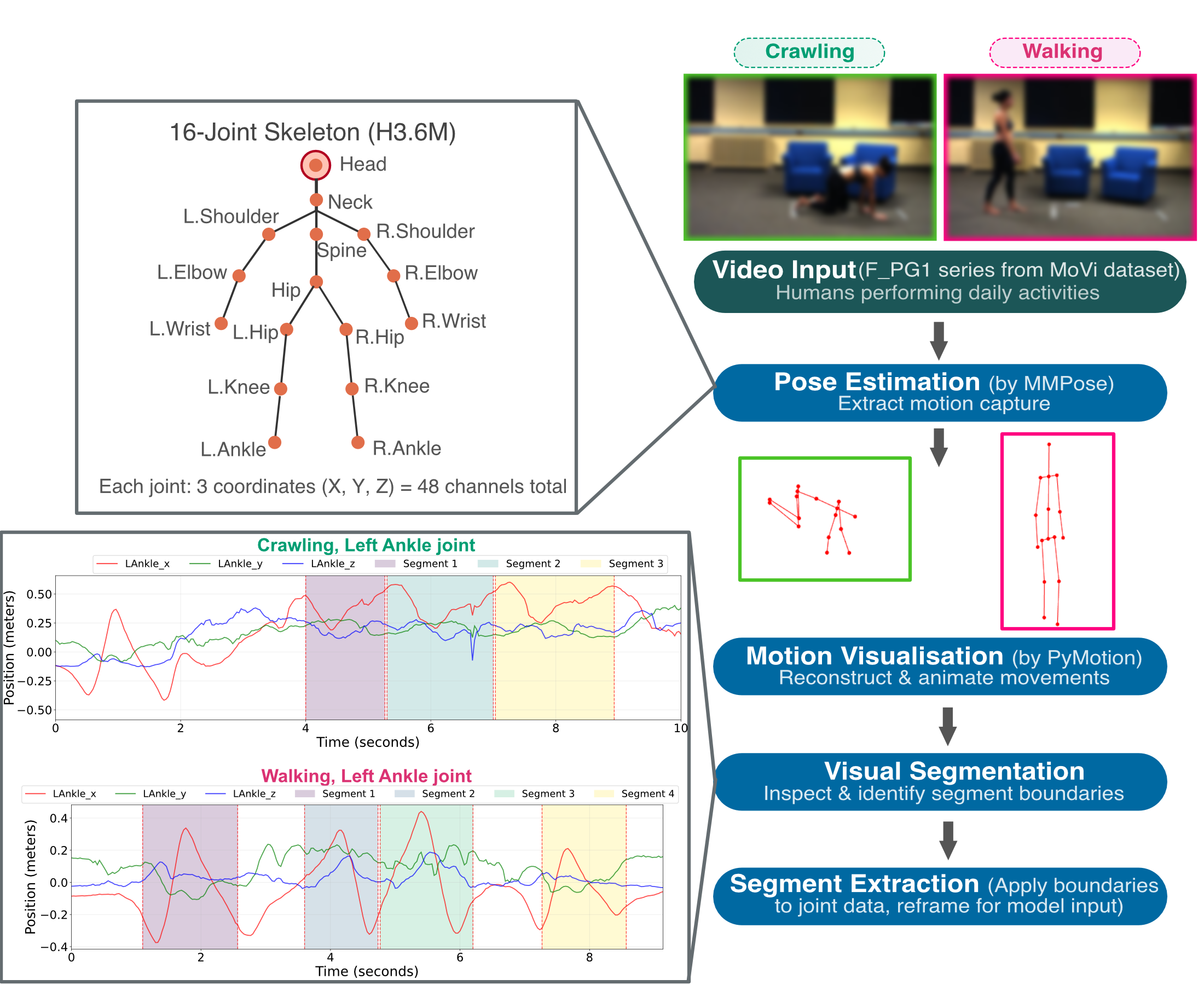}
\caption{\textbf{Data processing pipeline.}
Right: raw video frames from the \href{https://www.biomotionlab.ca/movi/}{MoVi dataset} (dataset is publicly available under a research license; all participants provided written informed consent for their data, including video footage, to be used by other researchers \cite{Ghorbani2020MoVi:Dataset}) illustrated for crawling 
and walking are passed through MMPose to extract 3D joint positions, 
then animated using the \href{https://github.com/UPC-ViRVIG/pymotion}{PyMotion library} \cite{Ponton2023SparsePoser:Data} for visual inspection. Each continuous 
recording is visually segmented by annotating the start and end frames 
of one complete movement cycle per activity. Left: the 16-joint 
skeleton model following the Human3.6M convention. Bottom-left: example joint-position 
time series for the left ankle during crawling and walking, with colored segments 
marking individual movement cycles identified through visual inspection. As shown here, different segments have different durations.}
\label{fig2}
\end{figure}

We used the MoVi dataset~\cite{Ghorbani2020MoVi:Dataset}, a large-scale human motion capture corpus recorded at 120\,Hz and down-sampled to 30\,Hz. The dataset contains recordings of 77 participants performing a range of daily activities, from which we selected 16 motion categories that are common and consistent among all participants (see one frame of crawling and walking videos in Fig~\ref{fig2}). We then used the MMPose library~\cite{mmpose2020} to extract 3D joint positions for 16 anatomical landmarks following the Human3.6M skeleton convention (Fig~\ref{fig2}), yielding 48 channels per frame (16 joints $\times$ 3 Cartesian coordinates). Fig~\ref{fig4} shows a representative skeleton pose from one segment of each 16 activity categories, illustrating the diversity of body configurations across the selected motions.

\subsubsection{Motion segmentation}

Each subject's recording consisted of one continuous video containing all motion categories performed in sequence. We first extracted the part corresponding to each motion label using the segmentation scripts provided with the MoVi dataset~\cite{Ghorbani2020MoVi:Dataset}. However, the raw segmentation required further refinement, as participants varied in how many times they repeated a given motion and transitions between consecutive activities sometimes corrupted the beginning or end of a segment. We therefore performed systematic visual inspection of each video, annotating the starting and ending frame of one complete movement cycle that was perceived as fully natural and coherent. We ensured in this process that the natural variability of performing one identical motion among participants was preserved. Fig~\ref{fig2} illustrates this process for the left ankle joint in two example activities (crawling and walking), showing the three Cartesian coordinates over time with individual movement cycles marked as colored segments. This ensured that all segments of a given motion category began and ended in a consistent posture, providing a reliable basis for cross-subject comparison. Segments shorter than 10 frames were discarded. This procedure yielded 1385 segments of variable length across all participants and motion categories.

\subsection{Feature extraction}

We compared three feature extraction strategies, each embodying different assumptions about movement structure. All three methods operate on the same set of segmented joint-position trajectories and produce a fixed-length feature vector for each segment, which is then passed to downstream evaluations. Crucially, we carry all three strategies through the entire analysis rather than selecting a single strategy, because each affords a distinct form of insight: Legendre coefficients allow the contribution of posture and dynamics to be cleanly separated by polynomial degree; TMP weights preserve the identity of each joint and coordinate, enabling joint-level interpretation and temporal reconstruction; and the autoencoder provides a data-driven baseline without assumptions against which the two structured methods can be compared.

\subsubsection{Temporal Movement Primitives (TMPs)}

Temporal Movement Primitives decompose each movement segment into a weighted sum of temporally smooth basis functions (primitives) learned under a Gaussian process prior~\cite{Endres2013ModelPrimitives,Leh2023AValidity}. Formally, a segment $\mathbf{x}(t)$ with $D = 48$ channels is modeled as
\begin{equation}
x_d(t) = \sum_{k=1}^{K} w_{dk}\,\varphi_k(t) + \varepsilon(t)
\end{equation}
where $\varphi_k(t)$ are $K$ shared basis functions (primitives) evaluated on a fixed grid of $T$ discrete time points, $w_{dk}$ are segment- and channel-specific weights, and $\varepsilon(t) \sim \mathcal{N}(0, 0.03^2)$ is Gaussian observation noise. The primitives are assigned a multivariate Gaussian prior with a radial basis function (RBF) covariance kernel, encouraging temporal smoothness. Weights receive independent standard Gaussian priors.

\begin{figure}[!htbp]
\centering
\includegraphics[width=\textwidth]{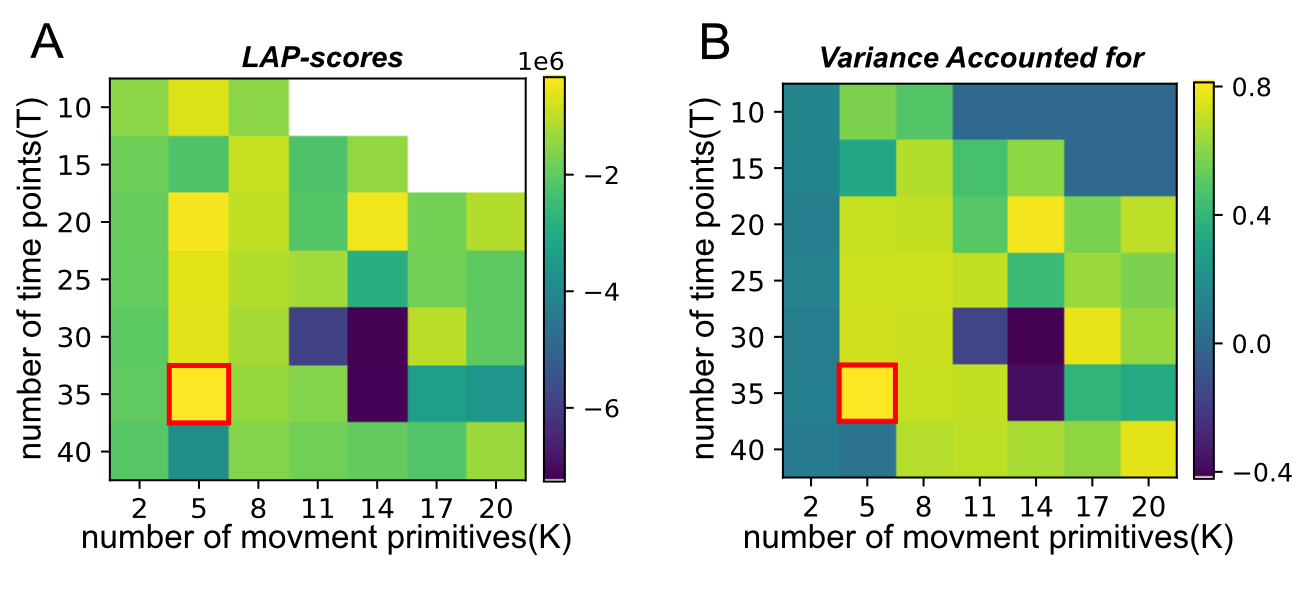}
\caption{\textbf{Joint hyperparameter selection for the TMP model.}
Heatmaps of (A) the Laplace approximation to the log marginal likelihood (LAP) and (B) the variance accounted for (VAF), evaluated over a grid of time-grid resolution $T \in \{10, 15, 20, 25, 30, 35, 40\}$ and number of primitives $K \in \{2, 5, 8, 11, 14, 17, 20\}$. For every $(T, K)$ cell, an independent TMP model was trained from scratch under the procedure described in the text (PCA initialization, 100 ADAM steps followed by 30 L-BFGS steps). The red square marks the LAP-maximizing configuration ($K = 5$, $T = 35$), which we adopted for all subsequent analyses. VAF saturates well before this optimum, indicating that further increases in $K$ or $T$ improve raw fit only marginally and are penalized by the marginal likelihood. White squares indicate infinite LAP values.}
\label{fig3}
\end{figure}

Model parameters (primitives and weights) were initialized via principal component analysis (PCA) on a random subset of 50\% of all segments (random seed 42), and then optimized in two stages: 100 iterations of ADAM with the default learning rate, followed by 30 iterations of L-BFGS for fine-tuning. Convergence was declared when the relative gradient norm fell below $5 \times 10^{-4}$ and the variance accounted for (VAF) exceeded 95\%. Following~\cite{Endres2013ModelPrimitives,Leh2023AValidity}, the two hyperparameters of the TMP model, the number of primitives $K$ and the number of time-grid points $T$, were selected jointly by Bayesian model comparison, using a Laplace approximation to the log marginal likelihood (LAP) computed via a block-wise Hessian, with the final VAF reported as a complementary fit-quality diagnostic. The joint LAP and VAF landscapes over the $(T, K)$ grid are shown in Fig~\ref{fig3}; we used $K = 5$ and $T = 35$ for all subsequent analyses. Importantly, the TMP weight space preserves the identity of each joint and coordinate, so that each weight can be traced back to a specific anatomical channel and primitive, making this representation suitable for interpretability analyses.

\subsubsection{Legendre polynomial coefficients}

As a second strategy, we projected each joint-position trajectory onto an orthogonal polynomial basis. Specifically, for each of the 48 channels within a segment, we normalized time to the unit interval $[0, 1]$ and fit shifted Legendre polynomials up to degree $M$ by ordinary least squares:
\begin{equation}
x_d(t) \approx \sum_{m=0}^{M} c_{dm}\,P_m(2t - 1)
\end{equation}
where $P_m$ denotes the $m$-th Legendre polynomial, evaluated via \texttt{scipy.special.eval\_legendre}. Each polynomial degree carries a distinct biomechanical interpretation: degree zero captures the mean posture (static configuration), degree one captures the linear trend (simplest temporal change), and degree two captures curvature (acceleration dynamics). The feature vector for each segment was formed by concatenating all coefficients across channels, yielding a dimensionality of $48 \times (M + 1)$.

We systematically varied $M$ from 0 to 9 to investigate how classification performance depends on the complexity of the temporal representation. As shown in Fig~\ref{fig6}A, the optimal polynomial degree was $M = 0$, which achieved the highest test accuracy with the smallest gap between training and test performance (indicating the best generalization). This produces 48-dimensional feature vectors, one coefficient per joint channel. As with TMPs, the Legendre representation preserves the identity of each joint and coordinate, making it directly interpretable in anatomical terms.

\subsubsection{Autoencoder latent embeddings}

The third strategy employed a feed-forward temporal autoencoder to learn compressed features through a reconstruction objective. The encoder applies a per-timestep linear projection ($48\to128$, LeakyReLU) followed by a masked average pool over the segment's valid (non-padded) time steps, and two fully connected layers ($128\to128$ with LeakyReLU and dropout = 0.2, then $128\to 32$) producing a 32-dimensional latent vector. The decoder applies two fully connected layers ( $32\to128\to128$, LeakyReLU and dropout = 0.2), broadcasts the resulting 128-dimensional vector across the original segment length, and projects each broadcast frame back to 48 dimensions through a final fully connected layer with Tanh activation. No recurrent, convolutional, or attention layers are used; temporal information is summarized by the encoder's mean pool and re-introduced by the decoder's broadcast. The model has approximately 54{,}000 trainable parameters.

Variable-length segments were zero-padded to the maximum length observed in the training set, and a binary mask tracked valid time steps. The network was trained to minimize masked mean squared error (computed only over non-padded positions) using the Adam optimizer (learning rate $10^{-3}$, weight decay $10^{-5}$) for up to 100 epochs. Training was halted early if the validation loss did not improve for 15 consecutive epochs. Data were split into training (70\%), validation (10\%), and test (20\%) partitions, stratified by motion label. The 32-dimensional latent vector served as the feature set for downstream classification. Unlike TMPs and Legendre coefficients, the autoencoder latent space does not preserve the identity of individual joint channels, which limits its interpretability and prevents channel-level analyses such as joint-specific feature selection or category-averaged reconstruction.

\subsubsection{Model comparison}

To place the three feature extraction strategies on a common information-theoretic footing, we compared them using Akaike's Information Criterion (AIC)~\cite{Akaike1974AIdentification,Burnham2004MultimodelSelection}, which trades off goodness-of-fit against representational complexity. For each strategy, we computed
\begin{equation}
\mathrm{AIC} = 2k - 2\ln\mathcal{L},
\end{equation}
where $k$ denotes the dimensionality of the feature vector supplied to the downstream classifier (240 for TMPs, 32 for the autoencoder, and 48 for Legendre coefficients), and $\ln\mathcal{L}$ is the cross-validated log-likelihood of the true class labels under a Random Forest classifier. We emphasize that $k$ here refers to the dimensionality of the features as a classifier input, not to the total number of free parameters internal to each feature extractor; this choice places the three strategies on a directly comparable axis of representational compactness and avoids an incommensurable mixture of per-segment fits (TMPs, Legendre) and globally shared network weights (autoencoder). The log-likelihood was estimated via 5-fold stratified cross-validation using the same folds as the classification analysis: within each fold, the Random Forest was trained on the training partition, and for every held-out sample the logarithm of the predicted probability assigned to its true class was accumulated, $\ln\mathcal{L} = \sum_{i=1}^{n}\ln P(y_i \mid \mathbf{x}_i)$. Because each sample appears in exactly one held-out fold, $\ln\mathcal{L}$ covers the entire dataset without optimistic bias. Following Burnham and Anderson~\cite{Burnham2004MultimodelSelection}, we compared strategies via AIC differences $\Delta_i = \mathrm{AIC}_i - \mathrm{AIC}_{\min}$, where $\mathrm{AIC}_{\min}$ is the lowest AIC among the three strategies. We applied the conventional evidence thresholds: $\Delta_i < 2$ indicates substantial support for strategy $i$ as a proper description of the data, $2 \leq \Delta_i < 4$ strong support, $4 \leq \Delta_i < 7$ considerably less support, and $\Delta_i > 10$ essentially no support relative to the best strategy. We additionally report the relative likelihoods $p_i = \exp(-\Delta_i/2)$, which quantify the probability that strategy $i$ would minimize AIC in a replicated experiment.

\subsection{Downstream evaluation}

All three feature extraction methods were evaluated through a common set of downstream analyses: movement classification, feature space analysis, and motion reconstruction.

\subsubsection{Movement classification}

Feature vectors from each method were split into training (75\%) and test (25\%) sets with stratified sampling (random seed 42). Training features were standardized to zero mean and unit variance using a StandardScaler fit on the training partition only; the same transformation was applied to the test set. To select an appropriate classifier, we compared four models: Random Forest (200 estimators), linear support vector machine (LinearSVC; $C = 1.0$, L2 penalty), multilayer perceptron (MLP), and logistic regression. As shown in \nameref{S1_Fig}, Random Forest consistently achieved the highest cross-validated accuracy across all three feature sets and was therefore used for all subsequent classification analyses. Five-fold stratified cross-validation was used to assess generalization, and final performance was reported on the held-out test set. Classification performance was quantified by accuracy, macro-averaged precision, recall, and F1-score, together with per-class confusion matrices normalized by row to show misclassification patterns.

\subsubsection{Feature space analysis}

To characterize the geometry of the learned feature spaces, we applied several complementary analyses spanning unsupervised dimensionality reduction, nonlinear visualization, supervised dimensionality reduction, posture-dynamics decomposition, and anatomical feature selection.

\paragraph{Dimensionality reduction and visualization.}
Principal component analysis (PCA) was used to quantify the intrinsic dimensionality of each feature space by examining the cumulative variance explained as a function of the number of components. We identified the number of components required to reach 90\% of the total variance as a summary measure of how compactly each representation concentrates its information. The first two principal components were also used to visualize the feature spaces in two dimensions, providing an initial assessment of category separation under linear projection.

Because PCA is blind to class labels, we additionally applied linear discriminant analysis (LDA) to quantify how much of the total variance in each feature space is relevant for discriminating between motion categories. LDA identifies the linear projections that maximize between-class separation relative to within-class scatter. By comparing the number of components required for 90\% of between-class variance (under LDA) with the number required for 90\% of total variance (under PCA), we assessed the degree to which each feature space contains non-discriminative variation. The first two LDA components were also used for two-dimensional visualization.

To explore nonlinear category structure not captured by linear methods, we applied $t$-distributed stochastic neighbor embedding ($t$-SNE; perplexity $= \min(30, n-1)$), which provided two-dimensional visualizations of cluster structure for each feature set.

\paragraph{Posture versus dynamics decomposition.}
To quantify the contribution of postural versus dynamic information to classification, we conducted a posture removal experiment. For each segment, we subtracted the temporal mean of each joint channel, effectively removing the time-averaged body configuration. We then re-fit Legendre polynomials to the mean-subtracted data across degrees 0 through 9 and re-evaluated classification accuracy at each degree. The gap in accuracy between original and mean-subtracted conditions quantifies how much classification relies on postural information. We additionally classified using the temporal means alone (posture-only features, equivalent to degree-zero coefficients only) to assess whether static body configuration without any dynamic information suffices for accurate classification.

\paragraph{Identification of discriminative joints.}
To identify the most discriminative anatomical landmarks, we applied L1-regularized classification (LinearSVC with L1 penalty) to the TMP weight space, which drives uninformative feature weights to zero and thereby selects a sparse subset of joints and coefficients carrying the most classification-relevant information. The regularization parameter $C$ was varied systematically, and the optimal value was selected based on the best cross-validated test accuracy. For the optimal classifier, we computed the importance of each feature as the mean absolute L1 coefficient across all classes, and ranked all 240 TMP features (16 joints $\times$ 3 coordinates $\times$ 5 primitives) accordingly. The top-ranked features were examined to identify which anatomical joints dominate the ranking.

To verify that the identified joints are both sufficient and necessary for classification, we performed two complementary experiments: an inclusion test, in which classification was repeated using only the first movement primitive weights of the top-ranked joints, and an exclusion test, in which all movement primitive weights of these joints were removed and classification was repeated using the remaining features.

\paragraph{Weight distribution analysis.}
To characterize how the selected joints encode activity-specific information, we analyzed the distribution of TMP weights across motion categories for the top-ranked channels identified by L1 selection. For each channel, we computed the deviation of each category's mean first movement primitive weight from the global mean across all categories, along with the standard deviation across segments within each category. This deviation-from-global-mean representation highlights which activities produce distinctively large or small weights on each channel relative to the population average, making the category-specific signatures more visually apparent than raw weight distributions. Narrow within-category standard deviations indicate that a weight is consistent across segments of the same motion, while broad distributions suggest high variability across participants or repetitions.

\subsubsection{Motion reconstruction}
\label{sec:methods:reconstruction}

To investigate whether features that support accurate classification also preserve the temporal structure needed for natural-looking movement, we evaluated reconstruction quality through two complementary pathways.

In the first pathway, we assessed whether category-level feature sets can generate recognizable movements. For each motion category, we averaged the feature vectors across all segments belonging to that category and reconstructed a single representative trajectory. For TMPs, the averaged weight vectors were multiplied by the learned primitives and resampled to the original temporal resolution. For Legendre coefficients (degree $M = 0$), the averaged zeroth-degree coefficients were evaluated at the original time points, producing a constant trajectory at the category-mean joint position for each channel. For the autoencoder, we were unable to perform an analogous procedure because the encoder does not preserve per-channel joint information in the latent space, preventing meaningful category-level averaging and decoding. The resulting category-averaged reconstructions were animated using the Pymotion library and visually inspected to assess whether each strategy produces perceptually meaningful movement sequences.

In the second pathway, we quantified the reconstruction fidelity of each strategy at the individual-segment level, using each segment's own learned features. For each of the $N$ segments in the dataset, we recomputed the reconstructed signal at its original temporal length $T_i$: TMP reconstructions were obtained by multiplying each segment's learned weight matrix with the shared movement primitives and resampling through the model's RBF kernel to $T_i$ frames; Legendre reconstructions were obtained by re-evaluating the fitted polynomial basis at $T_i$ uniformly spaced normalized time points; and autoencoder reconstructions were obtained by passing each segment through the trained encoder-decoder and cropping the decoder output to the segment's true (unpadded) length. This per-segment reconstruction protocol follows standard practice in the movement-primitive literature~\cite{DAvella2003CombinationsBehavior,Tresch2006MatrixSets,Chiovetto2013InvestigatingSynergies} and ensures that all three strategies are evaluated on the same inputs that subsequently feed the classifier.

For each segment we computed three quantitative reconstruction metrics spanning complementary aspects of fidelity. Variance Accounted For (VAF) measures the fraction of total variance in the original signal that is captured by the reconstruction, and is the field-standard metric for movement-primitive models~\cite{DAvella2003CombinationsBehavior}. We computed VAF over all $J$ channels and $T_i$ timesteps of each segment as
\begin{equation}
\mathrm{VAF}_i \;=\; 1 \;-\;
\frac{\sum_{j,t}\bigl(x_{i,j,t} - \hat{x}_{i,j,t}\bigr)^2}
     {\sum_{j,t}\bigl(x_{i,j,t} - \bar{x}_{i,j}\bigr)^2},
\label{eq:vaf}
\end{equation}
where $x_{i,j,t}$ and $\hat{x}_{i,j,t}$ denote the real and reconstructed value of channel $j$ at time $t$ of segment $i$, and $\bar{x}_{i,j}$ is the temporal mean of channel $j$ in segment $i$. VAF takes values in $(-\infty,\,1]$, with higher values indicating better reconstruction; a value of 0 indicates that the reconstruction captures no more variance than the temporal mean, and negative values indicate that the reconstruction error exceeds the original variance.

Mean Per-Joint Position Error (MPJPE) is the standard reconstruction metric in the motion-capture literature~\cite{Ionescu2014Human3.6M:Environments,Pavllo20183DTraining}. After reshaping the $J = 3K$ channels into $K$ three-dimensional joint coordinates, we computed
\begin{equation}
\mathrm{MPJPE}_i \;=\;
\frac{1}{K\,T_i}\sum_{k=1}^{K}\sum_{t=1}^{T_i}
\Bigl\| \mathbf{p}_{i,k,t} - \hat{\mathbf{p}}_{i,k,t} \Bigr\|_2,
\label{eq:mpjpe}
\end{equation}
where $\mathbf{p}_{i,k,t}\in\mathbb{R}^3$ is the real position of joint $k$ at time $t$ in segment $i$, and $\hat{\mathbf{p}}_{i,k,t}$ is its reconstruction. MPJPE is reported in the same units as the input data and decreases with better reconstruction. Because position-domain metrics are dominated by posture (the time-averaged joint configuration) and can therefore mask inaccuracies in the underlying dynamics, we additionally computed the velocity-domain root-mean-square error (vRMSE):
\begin{equation}
\mathrm{vRMSE}_i \;=\;
\sqrt{
\frac{1}{J\,T_i}\sum_{j,t}
\Bigl(\dot{x}_{i,j,t} - \dot{\hat{x}}_{i,j,t}\Bigr)^2
},
\label{eq:vrmse}
\end{equation}
where $\dot{x}$ and $\dot{\hat{x}}$ denote first-order numerical time derivatives (central differences via \texttt{numpy.gradient}) of the real and reconstructed signals, respectively. vRMSE specifically penalizes temporal smoothing and is therefore particularly informative for assessing whether a feature set preserves movement dynamics rather than only average posture.

For each strategy, we report the mean of each metric across all segments together with its standard deviation, broken down by motion class. All metrics are computed only on valid timesteps; for the autoencoder, this required masking out the zero-padded region of each fixed-length input batch before computing the per-segment errors. The same data preprocessing, segmentation, and train/test protocol used for the classification analysis was applied here, and all three strategies were evaluated on identical segments. Together, the two evaluation pathways allow us to investigate the dissociation between discriminative and generative adequacy: the first pathway tests whether category-level features produce recognizable movements, while the second quantifies the reconstruction fidelity of individual-segment features.
\subsection{Data availability}
The underlying data and analysis code are available in the Zenodo repository at \href{https://doi.org/10.5281/zenodo.20707925}{https://doi.org/10.5281/zenodo.20707925}.

\section{Results}

To address our central question, what features enable movement classification, we apply three strategies described above to 16 daily activities from the MoVi dataset and evaluate them through classification, feature space analysis, and motion reconstruction (Fig~\ref{fig1}). We begin by comparing classification performance across the three strategies, then use formal model comparison to resolve the apparent equivalence between TMPs and Legendre coefficients. We next examine the internal structure of each feature space to understand discriminative features that lead to classification, showing that Legendre feature sets use the static body configuration, how the body is positioned, for activity classification rather than how it moves over time. We then identify a compact subset of anatomical joints whose coordinate information alone suffices for accurate classification. Finally, we reveal a dissociation between classification accuracy and movement generation quality: strategies that excel at distinguishing activities do not necessarily produce temporal information of natural-looking reconstructed movements.

\subsection{Dataset and movement segmentation}

We analyzed 16 daily activities from the \href{https://www.biomotionlab.ca/movi/}{MoVi dataset}, selected to span 
both visually similar movements (e.g., walking, jogging, running in spot) 
and highly distinct ones (e.g., crawling, jumping jacks, cross-leg sitting). 
Each of the 77 participants performed activities in a single continuous 
recording, from which we extracted 3D joint positions for 16 anatomical 
landmarks and segmented the continuous stream into individual movement 
cycles (Fig~\ref{fig2}; see Methods). Because participants repeated each 
activity a variable number of times and transitions between activities 
sometimes corrupted segment boundaries, we visually inspected every 
recording and annotated the start and end of each complete, natural 
movement cycle. This procedure yielded 1385 segments of variable 
length across all subjects and activities, which formed the common input 
to all three feature extraction strategies. To illustrate the variability and structure of the segmented data, \nameref{S2_Fig} shows the left and right wrist trajectories across all three coordinates for one random segment of each activity.

\begin{figure}[!htbp]
\centering
\includegraphics[width=\textwidth]{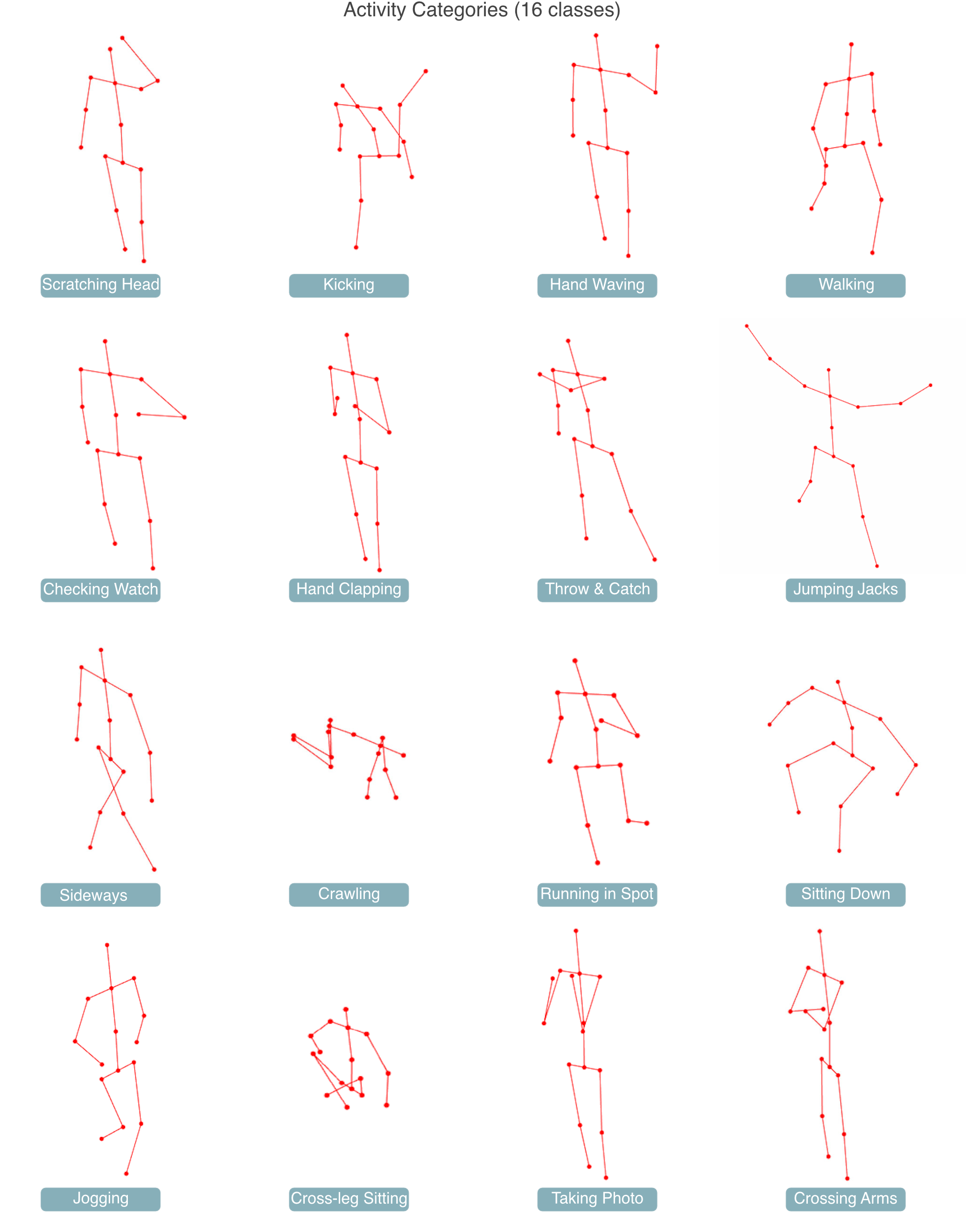}
\caption{\textbf{Representative keleton poses for each of the 16 daily activity categories.}
Each panel shows a single frame extracted from one movement segment, rendered using the 16-joint positions estimated by MMPose following the Human3.6M skeleton convention. The poses illustrate the diversity of body configurations across the selected motion categories, ranging from upright standing activities (walking, hand waving) to ground-level postures (crawling, cross-legged sitting). Note that many activities are visually distinguishable from their static body configuration alone, even without temporal information.}
\label{fig4}
\end{figure}

A first, qualitative observation motivates our central question. 
Fig~\ref{fig4} shows a single representative frame from each of the 16 
activities. Strikingly, most activities are immediately recognizable from 
this single static pose alone, without any temporal information: the 
ground-level sprawl of crawling, the wide limb spread of jumping jacks, 
the folded posture of cross-leg sitting, and the raised forearm of 
checking watch are each distinctive enough to identify the activity from 
one frame. This informal observation suggests that much of the information 
needed to distinguish between daily activities may be carried by the 
static configuration of the body rather than its temporal dynamics, a 
hypothesis that our subsequent analyses formalize and quantify.

\subsection{Motion classification}

We first compared four classifiers: Random Forest, linear SVM, multilayer perceptron, and logistic regression applied to the feature vectors extracted by each strategy. Random Forest consistently achieved the highest accuracy across all three feature sets (see \nameref{S1_Fig}) and was therefore adopted for all subsequent classification analyses.

\begin{figure}[!htbp]
\centering
\includegraphics[width=\textwidth]{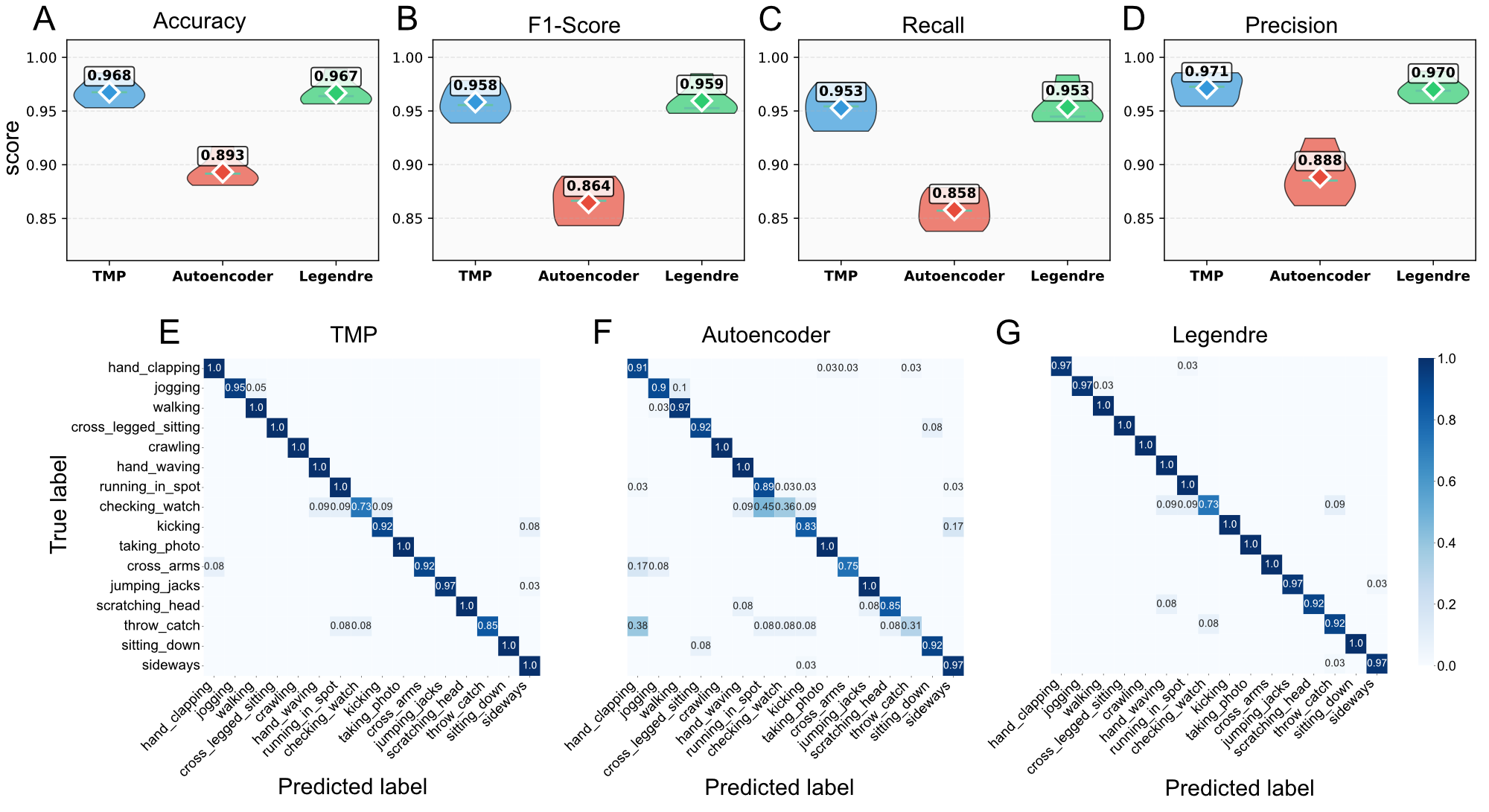}
\caption{{\bf Cross-validated classification performance and confusion matrix across three movement analysis strategies.} (A--D) Five-fold stratified cross-validation using Random Forest (200 estimators) on 16 motion categories. Violin plots show the distribution of scores across folds; diamond markers indicate the mean. (A) Accuracy. (B) Precision (macro-averaged). (C) Recall (macro-averaged). (D) F1-score (macro-averaged). (E--G) Row-normalized confusion matrices for each strategy: (E) Legendre polynomial coefficients (degree 0, 48 features), (F) TMP weights (5 primitives, 240 features), (G) autoencoder latent embeddings (32 features). Legendre and TMP achieve comparable classification accuracy, both substantially outperforming the autoencoder.}
\label{fig5}
\end{figure}

Fig~\ref{fig5}A-D shows the distribution of four classification metrics across five stratified cross-validation folds, and Fig~\ref{fig5}E-G shows the corresponding row-normalized confusion matrices. TMP weights and Legendre polynomial coefficients achieved remarkably similar performance. TMPs ($K = 5$ primitives, 240-dimensional feature vectors) yielded a mean accuracy of 96.8\%, precision of 97.1\%, recall of 95.3\%, and F1-score of 95.8\%. Legendre coefficients (degree $M = 0$, 48-dimensional feature vectors) achieved nearly identical values: accuracy of 96.7\%, precision of 95.9\%, recall of 95.3\%, and F1-score of 97.0\%. Both strategies exhibited minimal variance across folds, indicating stable generalization, and their confusion matrices (Fig~\ref{fig5}E,G) confirm strong diagonal dominance with only sporadic misclassifications between visually similar activities. The autoencoder latent embeddings (32-dimensional feature vectors) achieved notably lower performance: accuracy of 89.3\%, precision of 86.4\%, recall of 85.8\%, and F1-score of 88.8\%. Its confusion matrix (Fig~\ref{fig5}F) reveals more diffuse off-diagonal entries, suggesting that black box methods do not separate activity categories as cleanly as the other two feature sets.

The near-identical classification accuracy of TMPs and Legendre coefficients, despite their fundamentally different assumptions, raises an important question: if both strategies reach the same performance, what distinguishes them? Two observations are immediately relevant. First, Legendre coefficients achieve this performance with only 48 features (a five-fold reduction in dimensionality compared to the 240 TMP weights) suggesting that the Legendre basis captures discriminative information far more compactly. Second, and more strikingly, the optimal Legendre degree is $M = 0$, meaning that classification relies entirely on the zeroth-degree polynomial coefficient for each joint channel. Because the zeroth-degree Legendre coefficient captures the temporal mean of each trajectory, equivalent to the time-averaged position of each joints, this result indicates that the static body configuration alone accounts for nearly all of the between-activity discrimination. No contribution from higher-order temporal dynamics (linear trends, curvature, or more complex temporal patterns) is required to distinguish one daily activity from another at the category level. However, the question of whether this equivalence in classification accuracy also extends to representational parsimony and model fit requires a more formal comparison.

\subsection{Model comparison}

To resolve the apparent tie observed in classification accuracy, we compared all three strategies using Akaike's Information Criterion (AIC), which penalizes representational complexity while rewarding goodness-of-fit (see Methods for details). Table~\ref{table1} summarizes the results.

Legendre coefficients achieved the lowest AIC (1025.7), followed by the autoencoder ($\mathrm{AIC} = 1261.5$, $\Delta = 235.9$) and TMPs ($\mathrm{AIC} = 1455.9$, $\Delta = 430.2$). Following the evidence thresholds of Burnham and Anderson~\cite{Burnham2004MultimodelSelection}, both the autoencoder and TMPs received essentially no support ($\Delta > 10$) relative to Legendre. The relative likelihoods confirm this decisively: the probability that either alternative would minimize AIC in a replicated experiment is effectively zero ($p_i =3.85\times 10^{-94}$ for TMP and $6.09 \times 10^{-52}$ for autoencoder). Note that this comparison penalizes only the dimensionality of the feature space supplied to the classifier, not the total number of internal model parameters. If the total number of parameters were used instead, the autoencoder which contains substantially more trainable parameters($54k$) than either TMP(175 shared primitive parameters) or Legendre($0$), would rank even further behind.

Although TMPs matched Legendre in raw accuracy ($\sim$97\%), they required 240 features to do so. The AIC penalizes this increase in dimensionality, and the cross-validated log-likelihood of the TMP features ($\ln\mathcal{L} = -487.9$) was substantially lower than that of Legendre ($\ln\mathcal{L} = -464.8$), indicating that Legendre not only uses fewer features but also assigns higher confidence to correct classifications. The autoencoder, despite having the most compact feature set (32 dimensions), achieved the worst log-likelihood ($\ln\mathcal{L} = -598.8$), confirming that autoencoder feature sets are poorly suited for discrimination which is consistent with previous comparisons of handcrafted and deep features for activity recognition~\cite{Bento2022ComparingRecognition}.

\begin{table}
\centering
\caption{\textbf{AIC-based comparison of the three movement analysis strategies.} 
$k$ is the dimensionality of the feature vector supplied to the Random Forest classifier. $\ln\mathcal{L}$ is the cross-validated log-likelihood of the true class labels, summed across all held-out samples from 5-fold stratified cross-validation. AIC is computed as $2k - 2\ln\mathcal{L}$; lower values indicate a more parsimonious description of the data. $\Delta_i = \mathrm{AIC}_i - \mathrm{AIC}_{\min}$ quantifies evidence against each strategy relative to the best, and $p_i = \exp(-\Delta_i/2)$ gives the relative probability that strategy $i$ would minimize AIC in a replicated experiment. Evidence categories follow Burnham and Anderson~\cite{Burnham2004MultimodelSelection}.}

  \begin{tabular}{lrrrrrl}
    \toprule
    Strategy              & $k$ & $\ln\mathcal{L}$ & AIC    & $\Delta_i$ & $p_i$                  & Evidence                   \\
    \midrule
    Legendre coefficients & 48  & $-464.8$         & 1025.7 & 0.0        & 1.000                  & Substantial support (best) \\
    Autoencoder latents   & 32  & $-598.8$         & 1261.5 & 235.9      & $6.09\times10^{-52}$   & Essentially no support     \\
    TMP weights           & 240 & $-487.9$         & 1455.9 & 430.2      & $3.85\times10^{-94}$   & Essentially no support     \\
    \bottomrule
  \end{tabular}
\label{table1}
\end{table}

Beyond representational parsimony, the three strategies differ markedly in computational demand. Legendre 
coefficients require no shared training: each segment is fit in closed 
form by ordinary least squares in $0.27 \pm 0.01\,\mathrm{ms}$ on 
average, and the entire 1385-segment dataset is processed in 
$0.37\,\mathrm{s}$. TMPs jointly learn the $K=5$ shared movement 
primitives and the per-segment weights through 100 ADAM steps followed 
by 30 L-BFGS steps, requiring $41.6\,\mathrm{s}$ in total; once trained, 
a single segment can be reconstructed from its weights in 
$0.020 \pm 0.001\,\mathrm{ms}$. The autoencoder was trained for 100 
epochs of masked-MSE minimization, requiring 
$5217\,\mathrm{s}\;(\approx 87\,\mathrm{min})$ in total, with a per-segment 
encoder forward pass of $0.044 \pm 0.002\,\mathrm{ms}$. All timings were 
measured on a single CPU core of an Apple silicon (arm64) machine using 
PyTorch~2.9.0. Together, total fitting cost spans roughly five orders of 
magnitude across the three strategies: the autoencoder requires 
approximately $125 \times$ more training time than TMPs and 
$\sim 14{,}000 \times$ more than Legendre. At inference 
time, all three strategies produce a per-segment feature vector in well 
under one millisecond, indicating that the cost differential lies 
entirely in the one-time fitting/training phase rather than in 
deployment.

Taken together, the classification, AIC, and computational analyses converge on a clear conclusion: Legendre polynomial coefficients provide the most efficient discriminative feature set for motion classification, achieving top accuracy with the fewest dimensions, the highest classifier confidence, and by far the lowest computational cost. But what drives this efficiency? The fact that degree-zero coefficients (which encode only the time-averaged body configuration) suffice for near-perfect classification points to posture as the dominant discriminative feature. In the following section, we examine this hypothesis through feature space analysis and systematic decomposition of postural versus dynamic information.

\subsection{Feature interpretability}

\subsubsection{Posture as the dominant discriminative feature}

\begin{figure}[!htbp]
\centering
\includegraphics[scale=0.75]{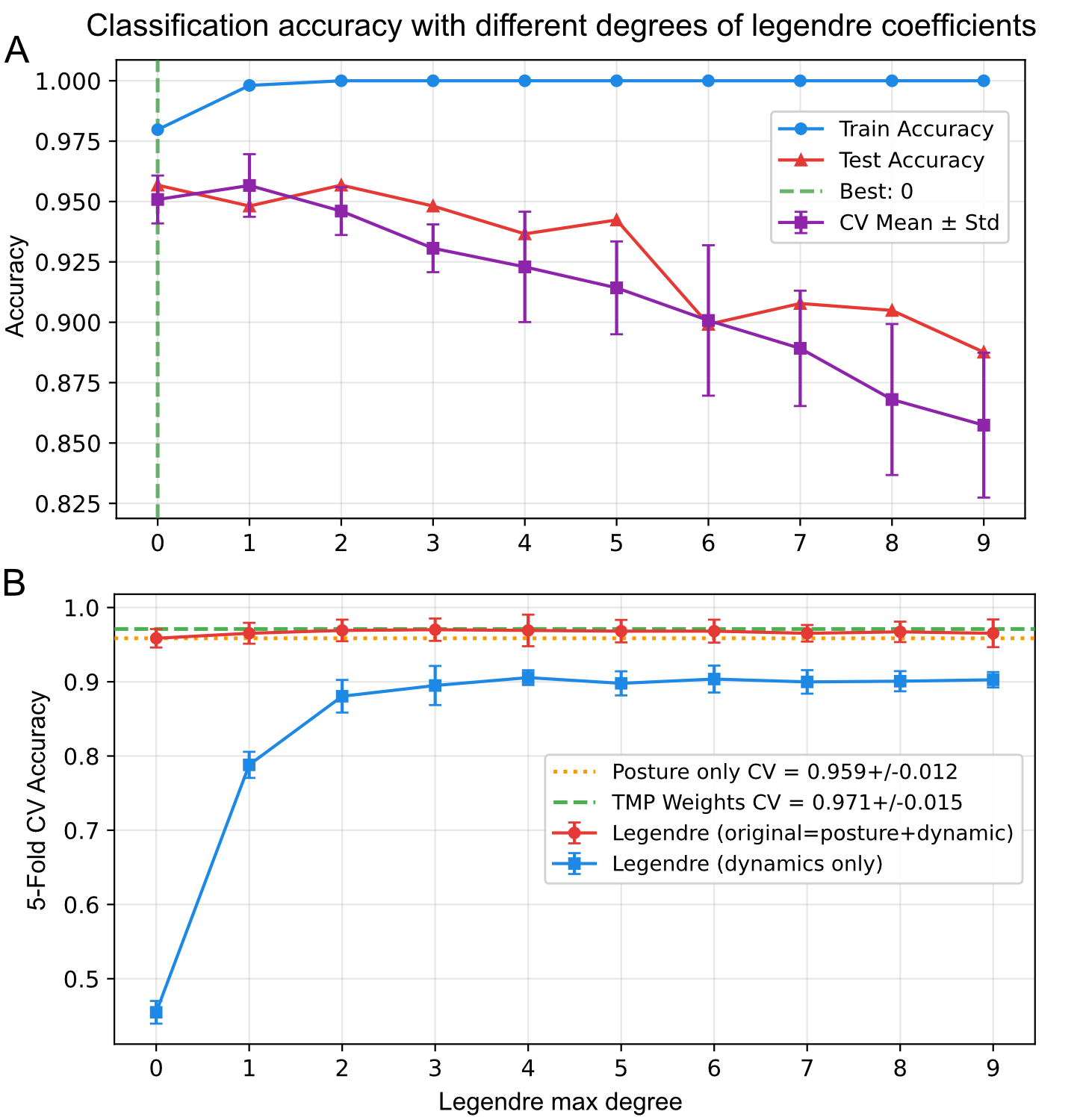}
\caption{\textbf{Performance of Legendre coefficients classification for different scenarios.}
(A) Legendre polynomial degree selection. Classification accuracy 
(train, test, and 5-fold cross-validated mean $\pm$ std) as a function 
of the maximum polynomial degree $M$. The optimal degree is $M = 0$ 
(dashed line), which achieves the highest test and cross-validated 
accuracy with the smallest train-test gap. Higher degrees add parameters 
without improving generalization. (B) Posture removal experiment. 
Red: Legendre classification on original data (posture + dynamics) 
across polynomial degrees 0--9. Blue: classification after subtracting the temporal mean from each individual segment, retaining only the dynamic component. Although dynamics-only accuracy remains well above chance (0.45 at degree 0 versus a chance level of
1/16=0.06), it falls behind the posture-only and original features, indicating that while inter-segment dynamics carry some discriminative information, removing category-averaged posture(temporal mean) severely degrades classification.
Horizontal reference lines show 
cross-validated accuracy for TMP weights (green dashed, 97.1\%) and 
posture-only features (orange dotted, 95.9\%). Removing posture causes 
a substantial drop in accuracy that higher-order dynamics cannot fully 
recover, confirming that static body configuration is the dominant 
discriminative feature.}
\label{fig6}
\end{figure}

The Legendre classification results revealed that degree-zero coefficients, which capture only the time-averaged body configuration, are sufficient for high-accuracy classification. As shown in Fig \ref{fig6}A, If we add more degrees of polynomials to Legendre coefficients fitting, this would add to the number of parameters without any enhancement in accuracy. This finding implies that postural configuration alone accounts for the majority of between-activity discrimination, with no contribution from temporal dynamics required at the category level.

To verify this, we performed a posture removal experiment. For each segment, we subtracted the temporal mean from each joint channel of all individual segments, which is the information captured by the degree-zero Legendre coefficient. We then re-fit Legendre polynomials to these mean-subtracted trajectories across degrees 0 through 9 and re-evaluated classification accuracy at each degree. As shown in Fig~\ref{fig6} B, classification accuracy dropped substantially after mean subtraction, confirming that the postural component carries the majority of discriminative information in Legendre classification. 
Notably, the dynamics-only information still classified well above chance, indicating that the temporal variation between segments does carry some activity-specific information. However, even when higher polynomial degrees were included to capture increasingly complex temporal dynamics, accuracy remained below that achieved by the only degree-zero coefficients of the original data. We additionally classified using the temporal means alone (posture-only features) and found that this minimal 48-dimensional description matched the performance of the full feature sets, providing further confirmation that posture dominates the discrimination.

This finding is also visually intuitive. Inspection of the representative skeleton poses for each activity category (Fig~\ref{fig4}) reveals that many daily activities adopt fundamentally different spatial arrangements of the limbs and joints and they are distinguishable from a single static frame, without any temporal information. The Legendre analysis formalizes this visual intuition: the zeroth-degree coefficients encode precisely these static configurations, and they suffice for classification.

\subsubsection*{Dimensionality and visualization of feature spaces}

Beyond classification accuracy, the internal geometry of each feature space can reveal how motion categories are organized relative to one another and how efficiently each feature set concentrates discriminative information. To characterize this geometry, we applied both unsupervised and supervised dimensionality reduction methods. We first used PCA to assess the total variance structure of each feature set, then applied linear discriminant analysis (LDA) to quantify how much of that variance is relevant for discriminating between motion categories.

\begin{figure}[!htbp]
\centering
\includegraphics[width=\textwidth]{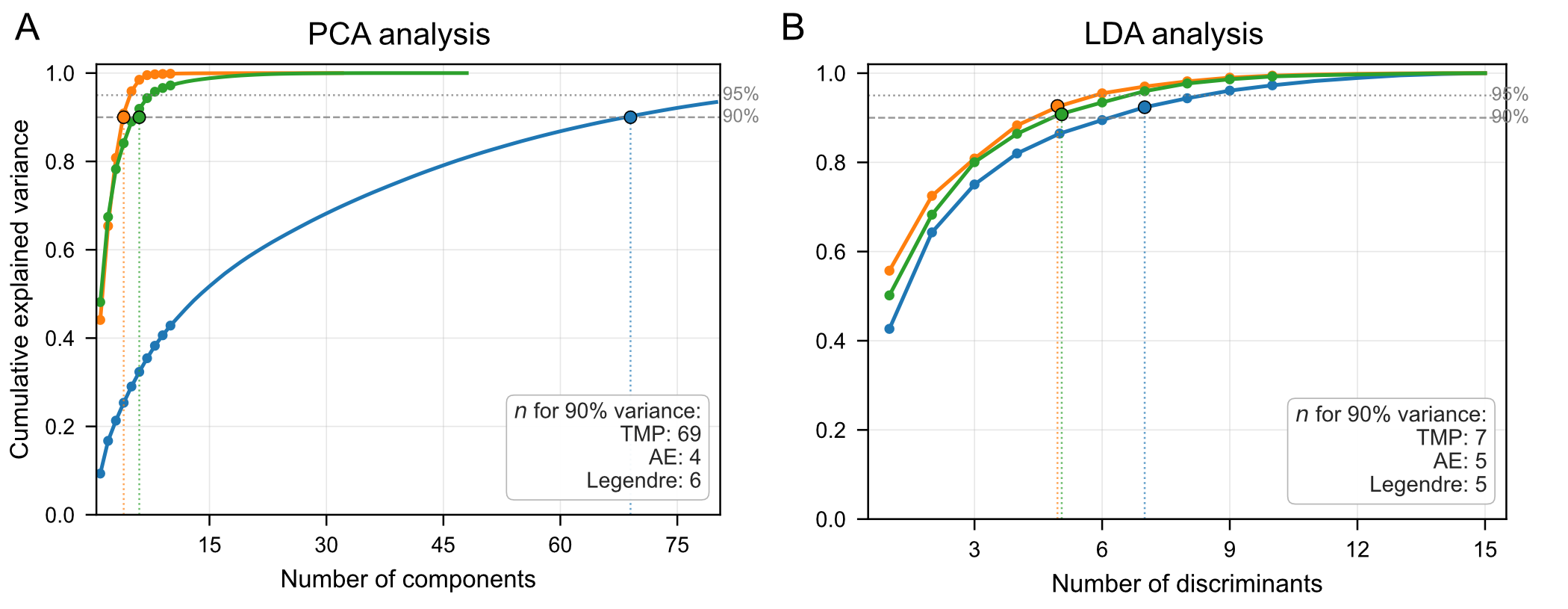}
\caption{{\bf Intrinsic dimensionality of the three feature spaces under unsupervised and supervised reduction.} (A) cumulative variance explained under PCA as a function of the number of principal components. Legendre and autoencoder features concentrate their variance in 6 and 4 components (90\% threshold), respectively, while TMP features require 69 components. (B) cumulative variance explained under LDA. The number of discriminant components for 90\% between-class variance drops to 5 for Legendre, 5 for the autoencoder, and 7 for TMPs. The ten-fold reduction for TMPs (69 to 7) indicates that most of the variance in the TMP feature space is within-class variation irrelevant to activity discrimination.}
\label{fig7}
\end{figure}

Fig~\ref{fig7}A (left panel) shows the cumulative variance explained under PCA as a function of the number of principal components. For Legendre coefficients and autoencoder latent embeddings, approximately 6 and 4 components suffice to capture 90\% of the total variance in our feature sets, respectively, indicating that the information in these feature sets is concentrated in a low-dimensional subspace. In contrast, the TMP feature space requires 69 components to reach the same threshold, consistent with its higher dimensionality (240 features) and suggesting that TMP weights distribute variance diffusely across many dimensions.

However, the critical question is not how many dimensions carry total variance, but how many carry \emph{discriminative} variance. Fig~\ref{fig7}B (right panel) shows the cumulative variance explained under LDA, which identifies the linear projections that maximize between-class separation relative to within-class scatter. Here, a striking asymmetry emerges. The number of discriminant components required for 90\% of the between-class variance is 5 for both Legendre and the autoencoder, essentially unchanged from PCA, but drops from 69 to just 7 for TMPs, a ten-fold reduction. This contrast reveals that the TMP feature space contains substantial within-class variance that is irrelevant to activity discrimination. PCA, which is blind to class labels, captures this uninformative variation alongside the discriminative signal, inflating the apparent dimensionality. LDA strips away the within-class scatter and shows that the discriminative structure of the TMP space is in fact comparably low-dimensional to Legendre and autoencoder feature sets. 

\begin{figure}[!htbp]
\centering
\includegraphics[width=\textwidth]{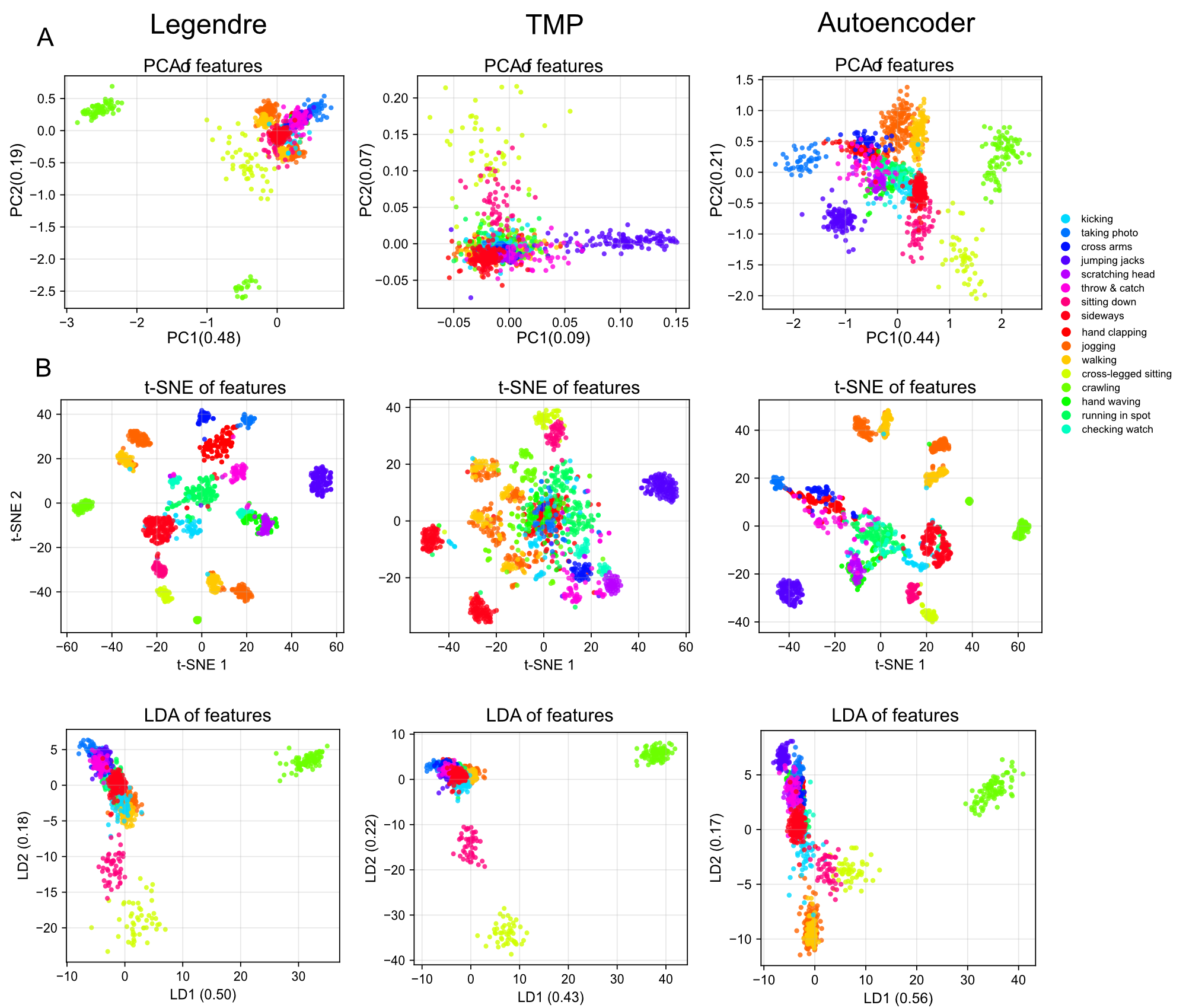}
\caption{{\bf Visualization of feature spaces under three dimensionality reduction methods.} Each column corresponds to one feature extraction strategy (left: Legendre polynomial coefficients, center: TMP weights, right: autoencoder latent embeddings). (A) projection onto the first two principal components. (B) $t$-SNE embeddings (perplexity $= \min(30, n-1)$). (C) projection onto the first two linear discriminant components. The Legendre feature space exhibits the cleanest category separation across all three visualization methods. Colors denote motion categories.}
\label{fig8}
\end{figure}

Fig~\ref{fig8} provides complementary visualizations of the three feature spaces under three dimensionality reduction methods: PCA (top row), $t$-SNE (middle row), and LDA (bottom row), each projected onto two dimensions. In the PCA projections (top row), the Legendre feature space shows partial separation, with activities that have highly distinctive body configurations (such as crawling and cross-legged sitting) separating clearly from the remaining categories, while activities that share similar upright postures (e.g., walking, scratching head, hand waving) cluster together. The TMP and autoencoder PCA projections show more diffuse distributions with greater overlap among categories. Projections onto higher PCA components (3rd and 4th) did not reveal additional discriminative structure, confirming that PCA alone cannot capture the full category separation in any of the three feature spaces.

The $t$-SNE embeddings (middle row; perplexity $= \min(30, n-1)$) reveal substantially more category structure through nonlinear dimensionality reduction. The Legendre $t$-SNE embedding exhibits the cleanest cluster separation, with nearly every activity category forming a distinct, compact island. The TMP $t$-SNE shows more congestion in the central region, with several categories overlapping. The autoencoder $t$-SNE falls between the two, with some well-separated clusters and some overlapping regions, reflecting its intermediate classification performance.

The LDA projections (bottom row) maximize class separability in two dimensions. The Legendre LDA projection reveals well-distributed clusters with good separation across the two-dimensional space, consistent with its compact and discriminatively efficient feature set. The TMP LDA projection shows a different pattern: a few categories form very tight, well-separated clusters, while many categories appear condensed into overlapping groups. The autoencoder LDA shows moderate separation with several identifiable clusters.

Overall, the dimensionality analyses converge on a consistent picture. The Legendre feature space is the most naturally organized for classification: its total variance and discriminative variance are closely aligned, its intrinsic dimensionality is low, and its two-dimensional projections under both linear and nonlinear methods reveal clean category separation. The TMP feature space achieves comparable classification accuracy but in much larger space of non-discriminative variation, requiring supervised methods (LDA) to recover its compact discriminative structure. The autoencoder feature space, despite its low total dimensionality, does not organize information as efficiently for discrimination, consistent with its lower classification performance and AIC ranking.

\subsubsection{Identification of discriminative joints}

While Legendre coefficients allowed us to cleanly isolate the postural component through polynomial degree, we now turn to the TMP strategy to ask a complementary question: which anatomical landmarks carry the most discriminative information? We focus on TMP weights here because, unlike the autoencoder latent space, they preserve the identity of each joint and coordinate, and unlike Legendre degree-zero coefficients (a single value per channel), they retain the full set of primitives per joint, making them the suitable for joint-level feature selection. We applied L1-regularized feature selection (LinearSVC with L1 penalty), which drives uninformative feature weights to zero and thereby identifies a sparse subset of joints critical for classification.

\begin{figure}[!htbp]
\centering
\includegraphics[width=\textwidth]{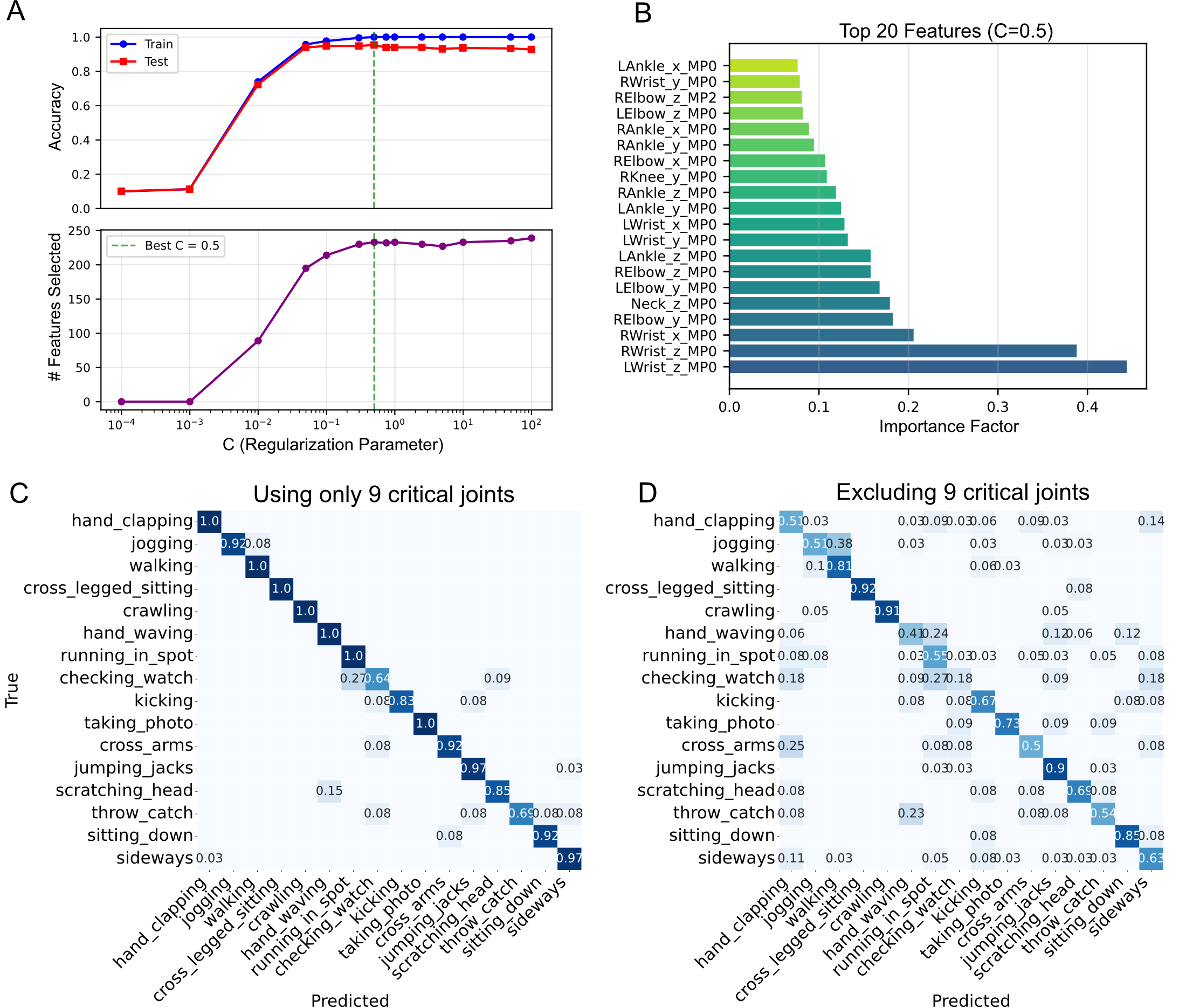}
\caption{{\bf L1-regularized feature selection.} (A) Upper panel: classification accuracy (train and test) as a function of the regularization parameter $C$ for L1-penalized LinearSVC applied to TMP weights; lower panel: number of non-zero features retained at each $C$ value. The dashed line marks the optimal $C = 0.5$. (B) Feature importance ranking: horizontal bars show the mean absolute L1 coefficient for each of the 20 top-ranked features, with coordinates of nine joints (wrists, elbows, neck, ankles, knees) dominating the ranking. (C) Row-normalized classification confusion matrix using only the first movement primitive weights of the top 9 joints, demonstrating that this compact subset preserves high classification accuracy. (D) Row-normalized confusion matrix after \emph{excluding} all movement primitive weights of the top 9 joints, showing a substantial drop in performance across nearly all categories. The contrast between (C) and (D) confirms that the identified joints carry discriminative information that is both sufficient and not recoverable from the remaining skeleton.}
\label{fig9}
\end{figure}

Fig~\ref{fig9}A (upper panel) shows classification accuracy as a function of the regularization parameter $C$, with both training and test accuracy plateauing at the optimal value of $C = 0.5$ (dashed line). Fig~\ref{fig9}A (lower panel) shows the corresponding number of non-zero features retained at each regularization level. we ranked all features (joints with their 3 cartesian coordinate) by their importance, defined as the mean absolute L1 coefficient across all classes. Across the 20 top-ranked features shown in Fig~\ref{fig9}B, different coordinates of nine joints consistently received the largest non-zero coefficients: both right and left joints of wrist, elbow, ankle, and knee, as well as the neck. Notably, these joints correspond to the endpoints and intermediate articulations of the limbs rather than the torso or pelvis, suggesting that the spatial positioning of the extremities is particularly informative for distinguishing between daily activities.

To test whether these 9 joints are not only the most important but also sufficient for classification, we restricted the TMP feature space to only the first movement primitive weights associated with these joints ($[4$ paired joints $\times 2 + 1$ neck joints] $\times 3$ coordinates $= 27$ features) and repeated the classification pipeline. Despite this reduction from the original 240 dimensions, this compact feature set achieved 96\% cross-validated test accuracy, statistically equivalent to the full TMP feature space. The confusion matrix for this reduced set (Fig~\ref{fig9}C) confirms strong diagonal dominance, with most activity categories classified at 0.92 or above. To further verify that these 9 joints are not merely redundant with information available elsewhere in the skeleton, we performed the complementary exclusion experiment: we removed all movement primitive weights of these 9 joints and classified using only the remaining 7 joints ($16-9$ joints $\times 3$ coordinates $\times 5$ primitives $= 105$ features). As shown in Fig~\ref{fig9}D, cross-validated test accuracy dropped to 80\%, with substantially more off-diagonal confusion across nearly all activity categories. This double dissociation confirms that the nine identified joints carry discriminative information that is both sufficient for high-accuracy classification and not recoverable from the remaining joints, even when including higher-order movement primitives.

\begin{figure}[!htbp]
\centering
\includegraphics[width=\textwidth]{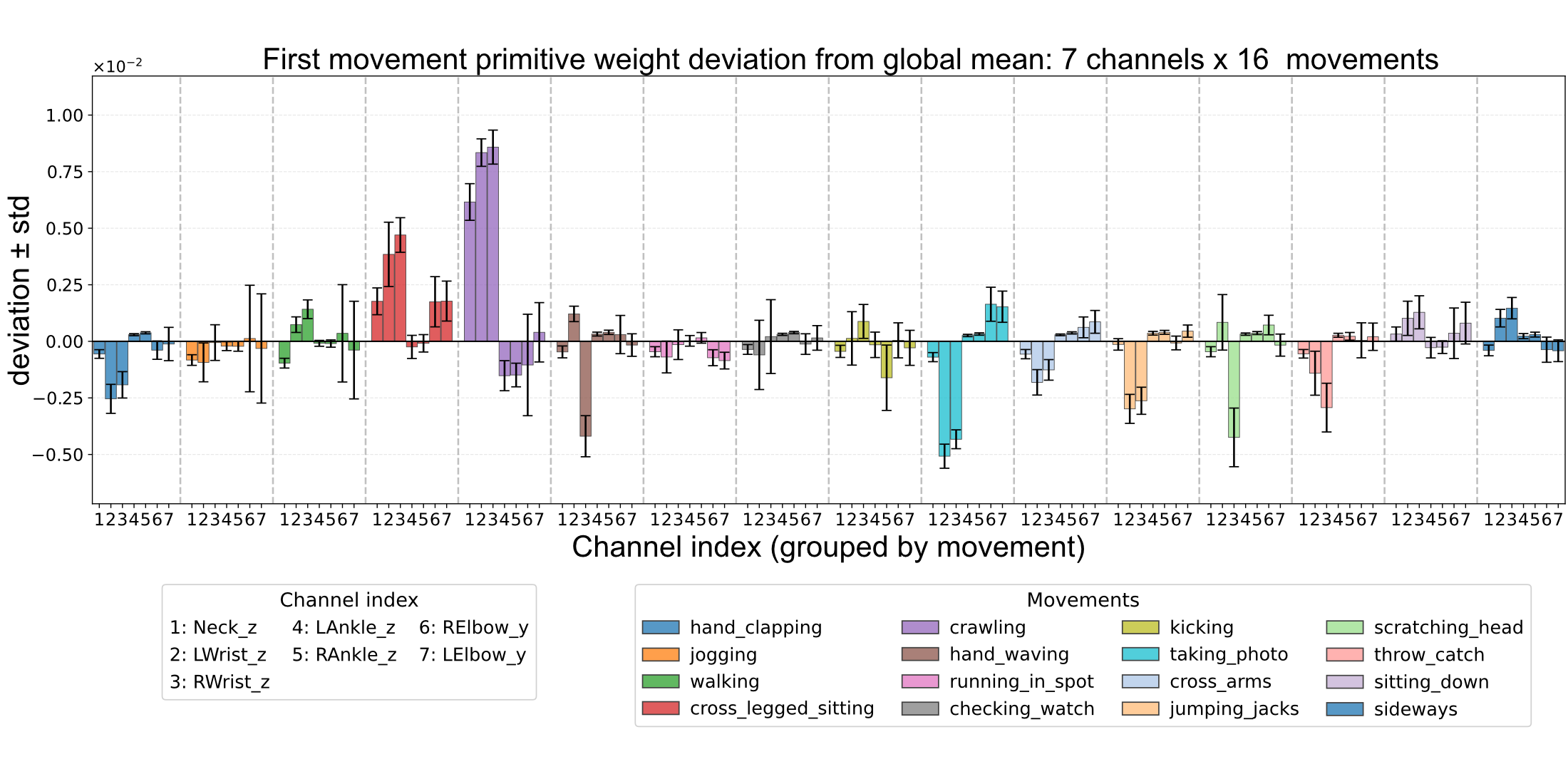}
\caption{{\bf Deviation of TMP weights from the global mean for top-ranked discriminative channels.} For each of the seven channels identified by L1 selection (Neck\_z, LWrist\_z, RWrist\_z, LAnkle\_z, RAnkle\_z, RElbow\_y, LElbow\_y), bars show the deviation of each motion category's mean first movement primitive weight from the global mean across all categories ($\pm$ standard deviation across segments). Positive deviations indicate that the weight for a given activity exceeds the cross-category average; negative deviations indicate it falls below. Each channel shows large deviations for a distinct subset of activities while remaining near zero for others, demonstrating that the nine selected joints contribute complementary discriminative information. Upper-limb channels (wrists, elbows) deviate most for arm-positioning activities, while lower-limb channels (ankles) deviate most for locomotor activities. Narrow error bars indicate high within-category consistency across participants.}
\label{fig10}
\end{figure}

To understand how these joints encode activity-specific information at the level of individual channels, we examined the distribution of first movement primitive weights across motion categories for the top-ranked features identified by L1 selection. To make the differences between categories more visually apparent, we computed the deviation of each category's mean weight from the global mean across all categories. Fig~\ref{fig10} shows these deviations ($\pm$ standard deviation across segments) for seven channels: Neck\_z, LWrist\_z, RWrist\_z, LAnkle\_z, RAnkle\_z, RElbow\_y, and LElbow\_y. A positive deviation indicates that the first movement primitive weight for a given activity is larger than the cross-category average, while a negative deviation indicates it is smaller; values near zero mean the activity does not differ from the cross-category average of that channel.
 
The deviation profiles reveal two key properties. First, each channel shows large deviations for a distinct subset of activities while remaining near zero for others, indicating that different joints contribute to discriminating different groups of movements. For example, the wrist Z-coordinate channels (LWrist\_z, RWrist\_z) show the largest deviations for taking photo, cross-legged sitting, and crawling, all activities where the wrists are displaced vertically from their neutral standing position. The ankle channels (LAnkle\_z, RAnkle\_z) show prominent deviations for kicking, and jumping jacks, reflecting the role of lower-limb displacement in locomotor and dynamic leg movements. 

Second, the narrow error bars for most activity-channel combinations indicate that the weight deviations are consistent across participants and repetitions, confirming that TMP weights for the selected joints encode stable, motion-specific information. The wider error bars observed for a few categories (e.g., checking watch on the wrist channels and kicking on the ankle channels) likely reflect between-subject variability in which hand or leg was used to perform the action.
 
Together, these observations reveal that the nine selected joints collectively tile the activity space: no single joint separates all 16 categories, but each contributes to discriminating a different subset of activities based on the biomechanical role of that body part in the movement. This complementary specialization explains why the compact feature set preserves classification accuracy while the exclusion of these joints devastates performance. This result also complements the Legendre posture analysis from a different angle. Where the Legendre decomposition showed that static body configuration is the dominant discriminative feature, the L1 analysis and weight distribution analysis reveal \emph{which parts of the body} contribute most to that discrimination and confirm that these joints carry stable, category-specific information. Together, these findings indicate that motion classification relies primarily on the spatial arrangement of a compact set of limb endpoints and articulations, a finding with practical implications for clinical movement assessment, where tracking a small number of anatomical landmarks is often more feasible than full-body motion capture~\cite{Stenum2024ClinicalChange,Uhlrich2023OpenCap:Videos}.

\subsection{Motion reconstruction}

The classification and feature space analyses above established that Legendre degree-zero coefficients extract posture information for high performance accuracy of classification among different activities. In contrast, TMP features encode a mixture of postural and dynamic information and mostly rely on 9 critical joints. We next asked whether these strategies differ in their capacity to preserve or generate the temporal structure of movement, evaluating reconstruction through two pathways: visual assessment of animations generated from category-averaged features, and quantitative evaluation of per-segment reconstruction fidelity.

\subsubsection{Category-averaged reconstruction: visual assessment}
 
To assess whether category-level features can generate recognizable movements, we averaged the feature vectors across all segments of a given motion category and reconstructed a single representative trajectory for each strategy. For TMPs, the category-averaged weights were multiplied by the learned primitives to produce a full temporal trajectory; for Legendre degree-zero coefficients, the averaged coefficients yield a static posture (a single constant position per joint channel). However, we were not able to use autoencoder features to reconstruct motions in this way, because the information of each channel is not preserved during the input output coding. The resulting reconstruction segments were animated using the Pymotion library similar to our original input segment data and visually inspected (shown in source code repository: \href{https://github.com/arefefarah/mp-movement-classifier}{https://github.com/arefefarah/mp-movement-classifier}.) 
 
TMP animations produced smooth, recognizable movement sequences: the temporal dynamics characteristic of each activity, the cyclic leg swing of walking, the arm-raising pattern of jumping jacks, the forward-and-back rocking of crawling, were clearly preserved in the category-averaged reconstructions. Although the averaged movements appeared somewhat smoother and less variable than individual instances, they were identifiable as the intended activity. Legendre degree-zero animations, by contrast, produced a frozen body in a static posture with no temporal variation. While this static configuration was visually consistent with the intended activity (e.g., a bent-forward posture for crawling, an upright posture for walking), it contained no movement whatsoever. This qualitative observation confirms that TMPs are the only strategy capable of generating temporally structured, recognizable movements from category-level features, while Legendre features, despite their classification superiority, encode no temporal information that can be recovered through animation.

\subsubsection{Per-segment reconstruction: trajectory fidelity and quantitative metrics}
 
\begin{figure}[!htbp]
\centering
\includegraphics[width=\textwidth]{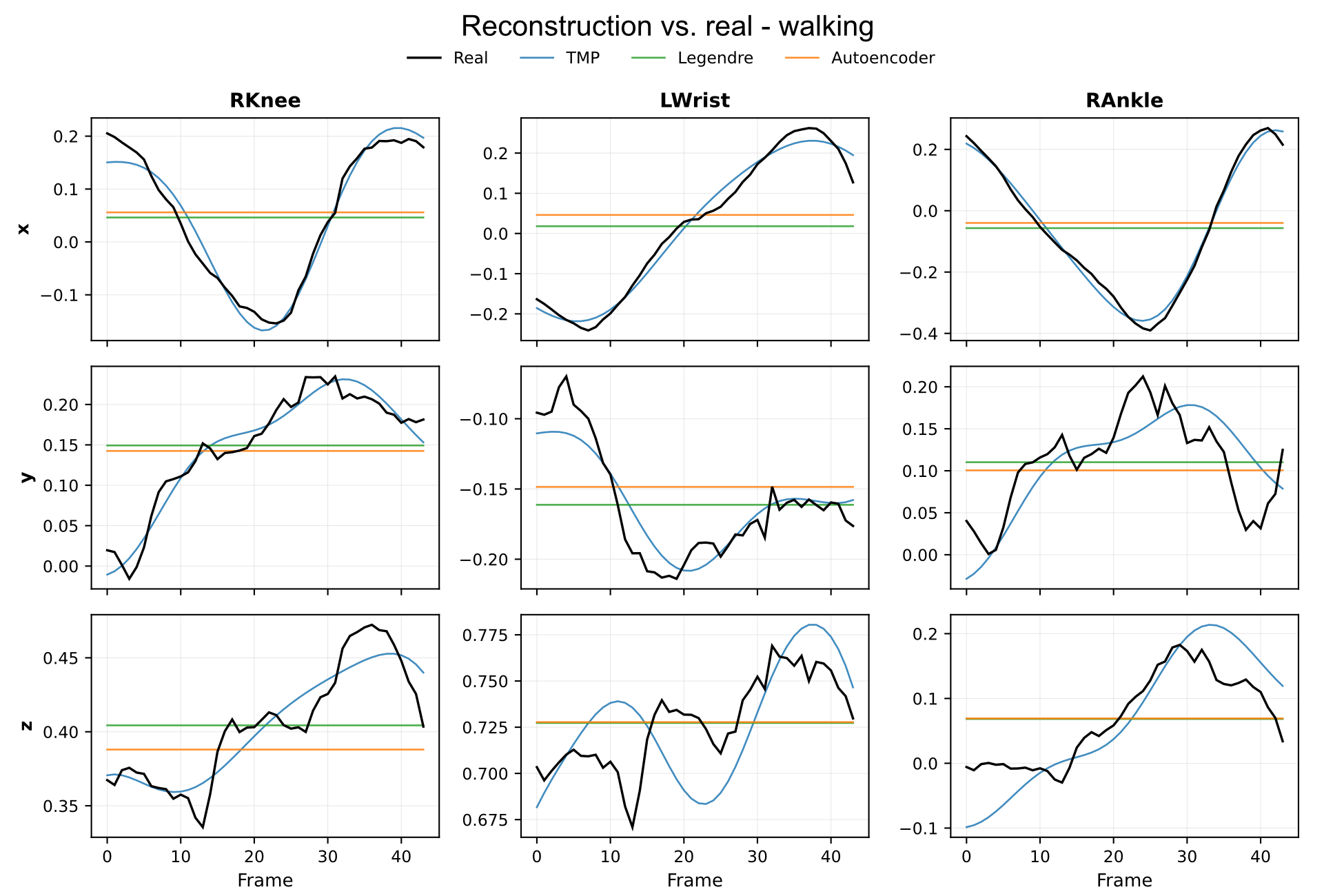}
\caption{{\bf Per-segment reconstructed versus original joint trajectories for walking.} Each panel shows one Cartesian coordinate ($x$, $y$, $z$; rows) for one joint (RKnee, LWrist, RAnkle) over the duration of a single walking segment, reconstructed from that segment's own learned features. Black: original trajectory. Blue: TMP reconstruction, producing smooth curves that capture the temporal dynamics. Green: Legendre degree-zero reconstruction, producing a constant value at the temporal mean of the channel. Orange: autoencoder reconstruction, also producing a flat output near the channel mean. TMP is the only strategy that preserves temporal structure at the individual-segment level.}
\label{fig11}
\end{figure}

\begin{figure}[!htbp]
\centering
\includegraphics[width=\textwidth]{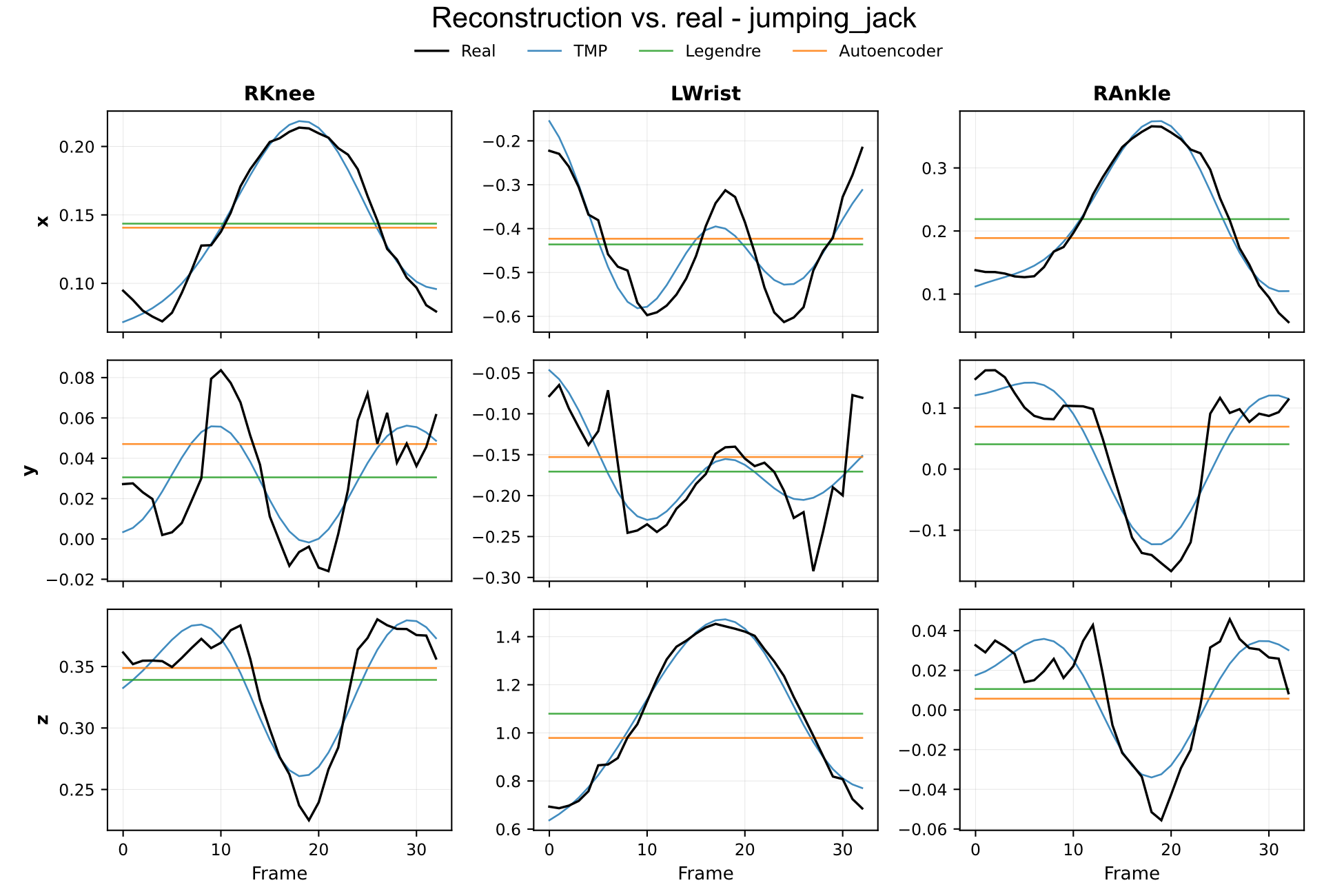}
\caption{{\bf Per-segment reconstructed versus original joint trajectories for jumping jacks.} Same format as Fig~\ref{fig11}. The large-amplitude oscillations characteristic of jumping jacks are captured by TMP reconstructions (blue) but are entirely absent from Legendre (green) and autoencoder (orange) reconstructions.}
\label{fig12}
\end{figure}

To complement the qualitative category-averaged assessment, we evaluated reconstruction fidelity at the individual-segment level, where each segment is reconstructed from its own learned features rather than from category averages. Fig~\ref{fig11} and Fig~\ref{fig12} show per-segment reconstructed trajectories for walking and jumping jacks, respectively, overlaid on the original trajectory for three discriminative joints (RKnee, LWrist, RAnkle) across all three Cartesian coordinates.

The contrast between strategies is immediately apparent. TMP reconstructions (blue) produce smooth temporal curves that track the overall shape of the real trajectories, capturing the oscillatory patterns characteristic of each activity. While the TMP curves do not reproduce high-frequency fluctuation in the original signal (consistent with the temporal smoothness imposed by the Gaussian process prior on the primitives), they preserve the fundamental dynamics: the cyclic knee and ankle excursions during walking, the large-amplitude wrist displacement during jumping jacks, and the correct phasing between joints. Legendre degree-zero reconstructions (green) produce flat horizontal lines at the temporal mean of each channel, capturing the average joint position but containing no temporal variation. Autoencoder reconstructions (orange) also produce flat outputs near the channel mean, indicating that the decoder fails to reconstruct the temporal dynamics of individual segments despite having been trained with a reconstruction objective.
 
\begin{figure}[!htbp]
\centering
\includegraphics{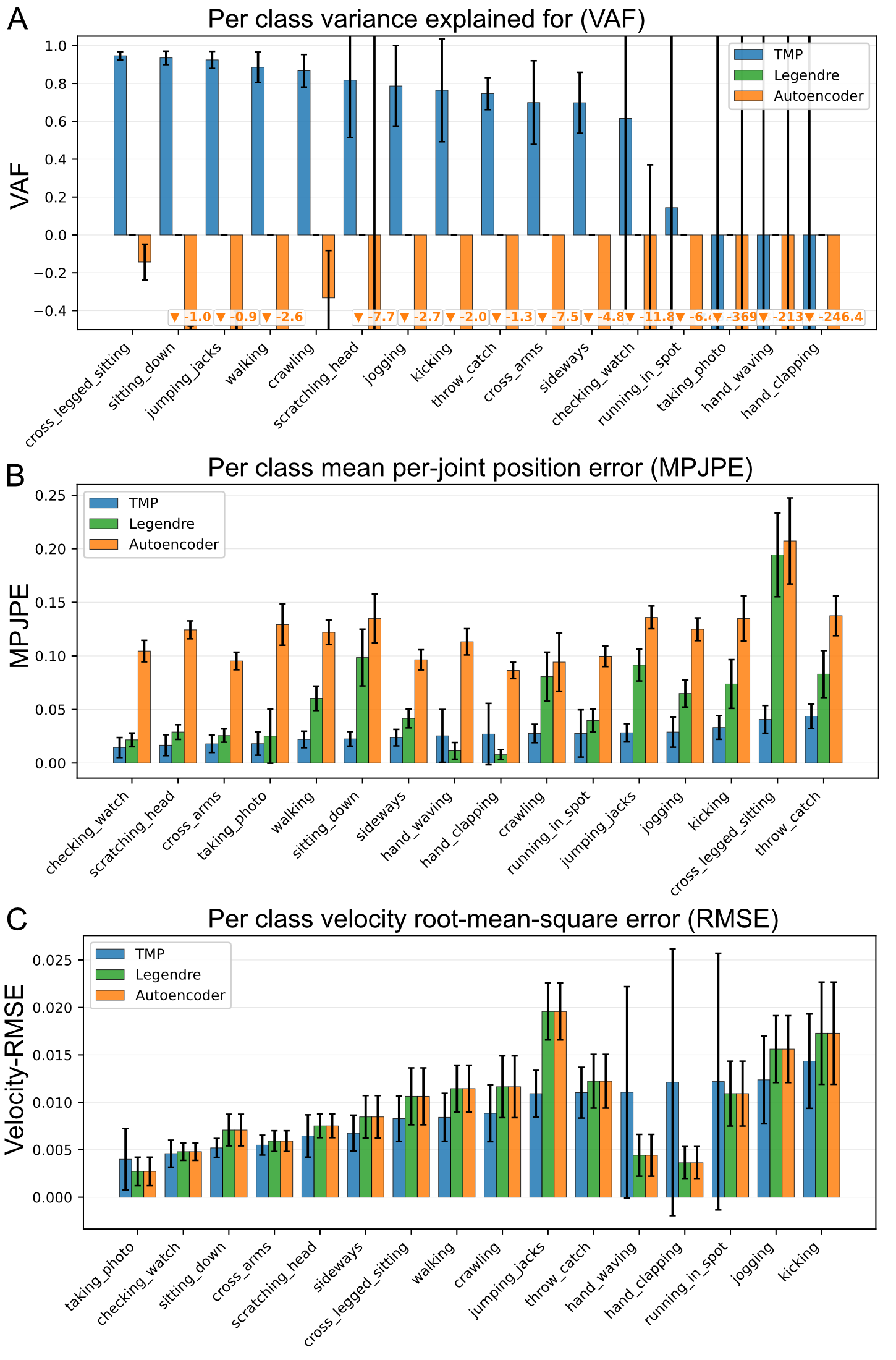}
\caption{{\bf Per-segment reconstruction metrics across 16 activity categories.} Each segment is reconstructed from its own learned features. (A) Variance accounted for (VAF). TMP reconstructions (blue) achieve positive VAF across all categories (0.15--0.95). Legendre degree-zero reconstructions (green) yield VAF $\approx 0$ across all categories (bars on the axis), as the degree-zero reconstruction equals the temporal mean by construction. Autoencoder reconstructions (orange) yield catastrophically negative VAF (off-scale values indicated by orange triangles with numerical labels). (B) Mean per-joint position error (MPJPE). TMP achieves the lowest error across all categories. (C) Velocity-RMSE. The three strategies are more comparable on this metric, with the largest differences for high-dynamic activities. Bars show the mean across segments; error bars denote $\pm$1 standard deviation.}
\label{fig13}
\end{figure}

\begin{figure}[!htbp]
\centering
\includegraphics[width=\textwidth]{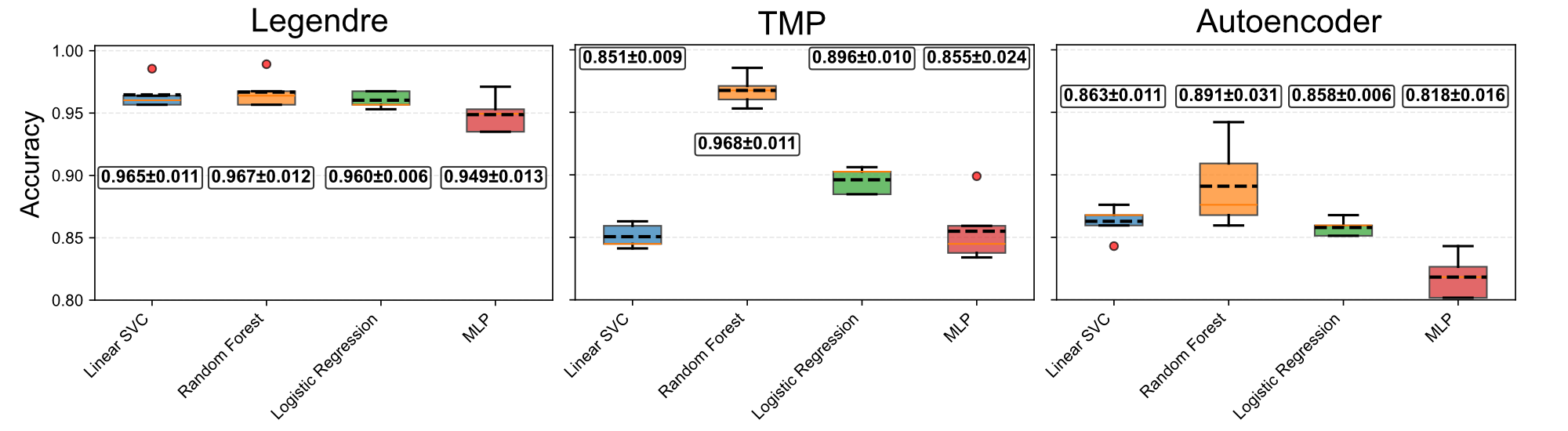}
\label{S1_Fig}
\end{figure}

\begin{figure}[!htbp]
\centering
\includegraphics[width=1\textwidth, height=0.95\textheight, keepaspectratio]{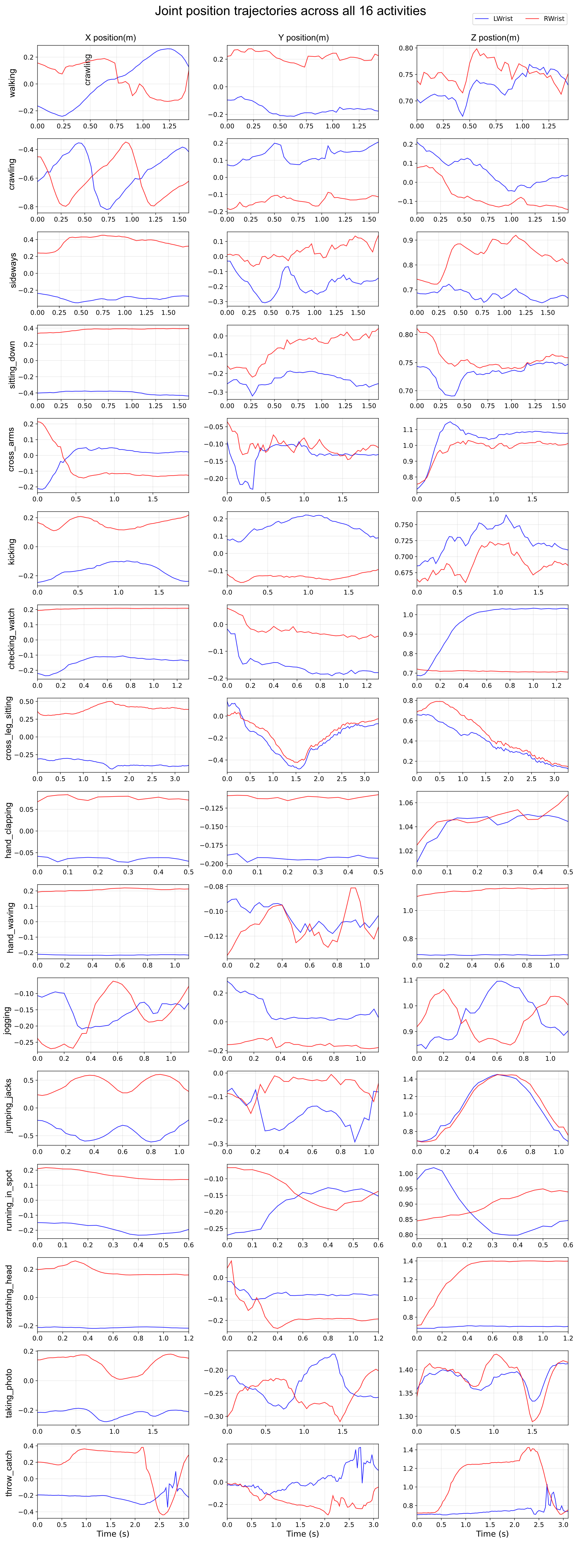}
\label{S2_Fig}
\end{figure}

We quantified these observations across all 16 activity categories using three complementary metrics (Fig~\ref{fig13}). TMP reconstructions achieved positive VAF across all 16 categories (Fig~\ref{fig13}A), ranging from approximately 0.15 (running in spot) to 0.95 (cross-legged sitting), indicating that TMP weights and their associated primitives preserve a substantial proportion of the temporal variance in the original trajectories. Activities with highly stereotyped, large-amplitude movements (cross-legged sitting, sitting down, jumping jacks) achieved the highest VAF values, while activities with more subtle dynamics (running in spot, checking watch) achieved lower but still positive values. Legendre degree-zero reconstructions achieved approximately zero VAF across all categories (green bars on the axis in Fig~\ref{fig13}A). This is expected given that the degree-zero coefficient for each channel approximates the temporal mean, so the reconstruction captures the average posture of each segment but none of its temporal variation. The autoencoder performed worst of all, with catastrophically negative VAF across every category.
This extreme negativity arises from two architectural 
properties of the minimal autoencoder. First, its decoder broadcasts a 
single latent-derived vector across all time steps, yielding a constant 
per-channel output (as with degree-zero Legendre). Second, its final 
Tanh activation bounds reconstructions to $[-1, 1]$, whereas the 
un-normalized joint positions extend beyond $1\,\mathrm{m}$; the 
resulting constant offset from each channel's true mean inflates the 
per-segment error most severely for low-amplitude activities, where the 
denominator of VAF (the signal's own temporal variance) is smallest. We 
retained un-normalized inputs so that all three feature-extraction 
strategies operated on identical joint-position data. Normalizing the 
inputs would raise the autoencoder's VAF toward zero by resolving this 
range mismatch, but would not yield positive VAF, because the broadcast 
decoder structurally cannot represent time-varying output. The 
autoencoder's poor reconstruction is therefore a property of its 
architecture and training objective rather than of preprocessing, 
consistent with our use of a deliberately minimal autoencoder as a 
proof-of-concept black-box baseline.

Fig~\ref{fig13}B shows the mean per-joint position error (MPJPE). TMP consistently achieved the lowest positional error across all 16 categories (approximately 0.015--0.04 in input coordinate units), confirming its superiority in spatial as well as temporal fidelity. Legendre MPJPE varied substantially across categories: it remained comparable to TMP for activities with small joint displacement from the mean posture (checking watch, cross arms) but rose sharply for activities involving large excursions (cross-legged sitting, kicking, throw and catch), where the static mean-position reconstruction deviates most from the actual trajectory at each time step. Autoencoder MPJPE was consistently the highest across nearly all categories, reflecting both poor temporal reconstruction and less accurate recovery of the spatial configuration.
 
Fig~\ref{fig13}C shows the velocity-RMSE, which isolates the fidelity of temporal dynamics by comparing frame-to-frame velocity profiles independently of static positional offset. Here the values for each strategy are more closed to each other overall, though TMP generally achieved the lowest values. For low-dynamic activities (checking watch, cross arms), all three strategies showed similar velocity-RMSE because the original movements themselves have limited temporal variation. The differences emerged most clearly for high-dynamic activities: jumping jacks showed elevated velocity-RMSE for Legendre an autoencoder reconstructions.

These results of recontructions reveal a clear dissociation between discriminative and generative adequacy of different methods. Legendre degree-zero coefficients achieved the highest classification accuracy and the best AIC, but their per-segment reconstructions capture zero temporal variance and produce only static postures when animated from category-averaged features. The TMP model which also achieved the highest accuracy required five times as many features and ranked last in AIC. But the TMP model is the only strategy that generated temporally structured, recognizable movements and achieved the highest per-segment reconstruction fidelity across all three metrics. The autoencoder performed worst on both fronts: lower classification accuracy than Legendre or TMPs, and the poorest reconstruction quality, with catastrophically negative VAF indicating that its decoder fails to faithfully reproduce even the individual segments it was trained on.
 
This analysis reflect a fundamental distinction between the information needed for classification versus reconstruction. Classification of daily activities relies primarily on the static body configuration, which is inherently low-dimensional and robust to averaging. Reconstruction requires the temporal evolution of joint positions over time. TMP features encode this temporal structure through weights on temporally smooth basis functions, and crucially, averaging weights across same motion segments preserves the temporal profile because averaged weights can still be multiplied by the learned primitives to generate smooth trajectories. Legendre degree-zero coefficients, by definition, encode no temporal information; their strength for classification is precisely their limitation for reconstruction.
The autoencoder's poor VAF despite acceptable training loss illustrates a further subtlety: minimizing mean squared reconstruction error is not equivalent to preserving temporal variance. Because joint position signals are dominated by their static mean, a network can achieve low MSE by accurately recovering the average posture while failing to capture the temporal fluctuations that VAF specifically measures. This observation reinforces the broader finding that features optimized for reconstruction fidelity (in the MSE sense) are not necessarily optimized for preserving the dynamic structure of movement.
 
This finding connects to the distinction between form and motion pathways in biological motion perception~\cite{Giese2003NeuralMovements,Vangeneugden2014DistinctDiscriminations}. Form (posture) suffices for recognizing \emph{what} activity is being performed, consistent with posture-template models~\cite{Lange2006ACues} and evidence that single-frame spatial information achieves substantial action recognition accuracy~\cite{Simonyan2014Two-StreamVideos}. Motion (temporal dynamics) is needed to perceive \emph{how} the movement unfolds and to generate reconstructions that appear natural~\cite{Knopp2019PredictingModels,Leh2023AValidity}. Our results formalize this dissociation computationally: no feature set optimally serves both classification and reconstruction, and the choice of feature strategy should be guided by the intended application.

\section{Discussion}

In this study, we compared three movement analysis strategies to determine which features enable the classification of daily activities and how classification relates to reconstruction fidelity. Our central finding is that the discriminative features for activity classification can be organized into two components: 1- The static body configuration (posture), which accounts for the majority of between-activity discrimination and achieve high performance classification; 2- The temporal displacement of a compact subset of nine joints (wrists, elbows, neck, ankles, and knees), which are most predictive joints for movement classification. Furthermore, we revealed a fundamental dissociation between classification and reconstruction: Legendre degree-zero coefficients achieved the best classification accuracy and model parsimony but produce no temporal dynamics, while TMPs were the only strategy capable of generating recognizable movements from category-averaged features. Below, we discuss the implications of these findings for biological motion perception, efficient computational modeling, and clinical movement analysis.

\subsection{Posture dominance and the form-versus-motion debate}

Our finding that static body configuration alone suffices for high-accuracy classification of daily activities aligns with a substantial evidence from the biological motion perception literature. Lange and Lappe\cite{Lange2006ACues} demonstrated that observers can recognize biological motion from configural form cues alone, proposing a template-matching model in which recognition proceeds through posture comparison without requiring temporal order. Critically, they found that spatially scrambling a point-light display which disrupt the spatial arrangement of dots while preserving their individual trajectories, devastated recognition performance. But temporally scrambling the frame order had comparatively little effect. This asymmetry between spatial and temporal scrambling maps directly onto our computational result: the Legendre degree-zero coefficients capture the spatial configuration and alone achieve 97\% classification accuracy. However, removing the postural component through mean subtraction causes accuracy to drop substantially even when higher-order temporal dynamics are included.
 
The scrambling literature provides further nuance that is consistent with our joint-selection findings. Troje and Westhoff~\cite{Troje2006TheDetector} showed that spatially scrambled displays retain some residual information about walking direction, attributable to the local motion of individual dots rather than global configuration. However, this residual direction discrimination is far weaker than intact recognition and disappears under vertical inversion, suggesting it relies on gravity-consistent motion cues from specific body parts (particularly the feet) rather than whole-body configuration. Our L1 analysis identified ankles and knees as two of the nine most discriminative joints, consistent with the special status of lower-limb motion in biological motion perception. At the same time, the dominance of whole-body postural configuration in our classification results supports the view that configural form cues, not local motion signals, carry the primary discriminative information for distinguishing between fundamentally different activities~\cite{Beintema2002PerceptionMotion,Lange2006ACues}.
 
It is important to note that the dominance of posture we observed may be partly attributed to our selection of 16 daily activities, some of which differ substantially in body configuration. For finer-grained discrimination within the same activity category such as distinguishing between emotional expressions in gait~\cite{Roether2009CriticalGait}, identifying individuals from their walking pattern~\cite{Troje2002DecomposingPatterns}, or detecting subtle pathological deviations in clinical populations, temporal dynamics likely become the critical discriminative feature. Previous work has shown that dynamics are more important than posture for such within-category discriminations~\cite{Troje2002DecomposingPatterns,Knopp2019PredictingModels}. Knopp et al.~\cite{Knopp2019PredictingModels} found that dynamics evidence lower bound (ELBO) scores were more predictive of perceived naturalness than pose ELBO scores in a Graphics Turing Test, suggesting that even when posture suffices for category-level recognition, dynamics govern the perception of movement quality and naturalness. Our reconstruction results are fully consistent with this view: posture sufficed for classification, but only TMPs, which explicitly encode temporal dynamics through smooth basis functions, could reconstruct natural-looking movement.

The dominance of low-dimensional postural information also offers a computational account of the remarkable speed of biological action recognition. Human and animal observers can categorize complex visual stimuli from extremely brief exposures, often within 150 ms and before extended temporal integration is possible~\cite{Thorpe1996SpeedSystem,Fabre-Thorpe2001AScenes}. Our finding that the static spatial configuration of a small set of joints already separates activities suggests that the visual system can exploit a compact, near-instantaneously available postural code: a single well-resolved pose may carry most of the information needed to recognize what action is being performed. This is consistent with hierarchical models of biological motion perception in which an initial stage matches each static frame against a set of stored posture templates, a stage already sufficient for form-based discriminations such as identifying the figure or its facing direction, before any analysis of how posture changes over time~\cite{Lange2006ACues,Theusner2014ActionSpace}. Temporal integration would then be reserved for the finer, within-category discriminations, such as identity, emotion, or pathological deviation, where dynamics become indispensable~\cite{Troje2002DecomposingPatterns,Roether2009CriticalGait}. This division of labor, fast posture-based categorization followed by slower dynamics-based refinement, fits naturally with the dual form and motion pathways of biological motion perception~\cite{Lange2006ACues,Giese2003NeuralMovements} and may reflect an efficient strategy for allocating limited neural processing resources.
 
\subsection{Less is more: feature efficiency}
 
A striking aspect of our results is that 48 features — the zeroth-degree Legendre coefficients for 16 joints in three coordinates — achieved classification accuracy comparable to the full 240-dimensional TMP feature space and superior to the 32-dimensional autoencoder latent embeddings. This extreme compactness echoes a broader trend in computational modeling: that carefully chosen, interpretable features can match or outperform high-dimensional learned feature sets for many practical tasks.
 
This principle has been demonstrated across machine learning. In activity recognition, Bento et al.~\cite{Bento2022ComparingRecognition} showed that handcrafted features generalize better than deep neural network feature sets for domain generalization, suggesting that explicit structural knowledge about the task can compensate for the representational capacity of large models. More broadly, the recent emergence of small and efficient models in artificial intelligence has underscored that model size and feature dimensionality are not the primary determinants of performance when the underlying task structure is well matched to the feature inputs. Small language models with only a few billion parameters now achieve competitive performance with models orders of magnitude larger on many benchmarks\cite{Abdin2024Phi-3Phone,Jiang2023Mistral7B}, and the field is increasingly recognizing that compact, well-structured representations can be more robust, interpretable, and resource-efficient than large, opaque ones.
 
Our findings instantiate this principle in the domain of human movement analysis. The Legendre degree-zero feature set is not merely compact; it is maximally interpretable, with each feature corresponding to the mean position of a specific anatomical joint in a specific coordinate. This one-to-one mapping between features and biomechanical quantities is precisely the kind of interpretability that has been argued to be essential for high-stakes applications in clinical and scientific contexts~\cite{Rudin2019StopInstead,Xiang2025ExplainableReview,Slijepcevic2023ExplainablePalsy}. Our AIC analysis confirmed that interpretable Legendre degree-zero features are not only compact but also statistically preferred: Legendre achieved the lowest AIC by a wide margin, indicating that it provides the most parsimonious account of the data relative to its dimensionality.
 
Furthermore, the L1 joint-selection analysis showed that, the discriminative information is concentrated in a subset of nine joints, reducing the effective dimensionality to 27 features (nine joints times three coordinates) for the first TMP primitive alone. This progressive compression, from 240 (full TMP) to 48 (Legendre degree-zero) to 27 (selected joints), suggests that the information required for activity classification has a remarkably low intrinsic dimensionality, consistent with Bernstein's~\cite{Bernstein67CoordinateMovements} observation that complex movements are organized through a reduction in the effective degrees of freedom. The practical implication is that we do not need to track the full body skeleton in clinical movement assessment systems: monitoring the 3D positions of a few key joints may suffice for reliable activity classification, substantially reducing the hardware and computational requirements of markerless pose estimation systems~\cite{Stenum2024ClinicalChange,Uhlrich2023OpenCap:Videos}. This is especially valuable for resource-constrained deployments such as continuous fall detection or activity monitoring in home and clinical settings, where tracking a small set of joints is far more practical than full-body motion capture.

\subsection{The classification-reconstruction dissociation and its implications}
 
The dissociation between classification accuracy and reconstruction quality is one of the conceptually important findings of this study, because it reveals that discriminative and generative adequacy impose fundamentally different requirements on the feature sets. Legendre degree-zero coefficients, which are optimal for classification, encode no temporal information and produce static reconstructions. TMPs, which are optimal for reconstruction, distribute discriminative information across a much larger feature space that is penalized by parsimony metrics. No single method optimally serves both purposes.
 
This dissociation has direct practical implications for the choice of movement features in different applications. For screening and classification tasks, such as identifying activity types in free-living monitoring, Legendre-type postural features offer the best combination of accuracy, compactness, and interpretability. For applications that require generating or reconstructing temporal movement sequences such as animation, rehabilitation biofeedback, or perceptual studies of movement naturalness, TMPs are the appropriate choice, as they are the only strategy that preserves temporal dynamics under feature averaging and produces recognizable movements~\cite{Knopp2019PredictingModels,Leh2023AValidity}.

\subsection{Limitations and future directions}
 
Several limitations should be considered when interpreting our results. First, our analysis focused on 16 daily activities that differ substantially in overall body configuration. As discussed above, the dominance of posture may be less pronounced for finer-grained discriminations within a single activity type (e.g., distinguishing healthy from pathological gait, or emotional from neutral walking). Future work should examine whether the Legendre-TMP dissociation holds for within-activity discriminations where temporal dynamics are expected to play a larger role. This is particularly relevant for clinical applications, where the diagnostically meaningful differences often lie in how a movement unfolds rather than in the overall posture: distinguishing a limping or antalgic gait from a healthy one, detecting the reduced arm swing and shuffling steps of Parkinsonian gait, or tracking subtle recovery of movement quality during rehabilitation. In such cases, the postural features that suffice for coarse activity classification may be inadequate, and the temporal dynamics captured by TMPs are likely to become the more informative representation. Relatedly, our framework could be used to probe the temporal dynamics of recognition itself, by quantifying how classification accuracy scales with the fraction of a movement cycle available, and comparing this to the minimum exposure human observers need for reliable action categorization.
 
Second, our autoencoder architecture was a feed-forward temporal autoencoder that summarizes a segment's time course through a single masked mean-pool in the encoder and reconstructs through a constant-broadcast decoder, with no recurrent, convolutional, or attention component. We adopted this minimal feed-forward architecture as a proof-of-concept benchmark: it has the fewest moving parts of any black-box method, while remaining sufficient to test whether any unsupervised deep representation can match interpretable parametric strategies on motion classification and reconstruction. More sophisticated architectures — for example, an LSTM-based recurrent autoencoder, a variational autoencoder with explicit temporal disentanglement~\cite{Noseworthy2020Task-ConditionedPrimitives}, or a graph convolutional autoencoder that respects skeletal topology\cite{Yan2019AData} — might achieve better performance on both classification and reconstruction. However, the interpretability advantage of Legendre and TMP features over any black-box method would remain.
 
Finally, our study analyzed individual movement segments rather than continuous activity sequences. Real-world movement monitoring involves transitions between activities, overlapping actions, and variable durations, all of which pose additional challenges for both classification and reconstruction that our segment-based analysis does not address.

\section{Supporting information}
\paragraph*{Fig S1.}
\textbf{Classifiers Comparison} comparing all 4 different classifier on 3 different feature set. Random forest classifier achieved best performance. 

\paragraph*{Fig S2.}
\textbf{Left and right wrist position trajectories across all 16 activities.} Each row shows one activity; columns show the X, Y, and Z coordinates (in metres) over the duration of one random segment. Blue: left wrist (LWrist); red: right wrist (RWrist). The trajectories illustrate the diversity of both static offsets (mean position) and temporal dynamics across activities, and the frequent left-right asymmetry within a single movement.
\clearpage

\section{Acknowledgments}
The authors would like to express sincere gratitude to Prof. Dominic Andres for his invaluable guidance throughout the Temporal Movement Primitives model implementation in this work.
The research was undertaken thanks in part to funding from the Connected Minds Program, supported by Canada First Research Excellence Fund, Grant \#CFREF-2022-00010. Also, the work supported by the Natural Sciences and Engineering Research Council of Canada and the Canada Foundation for Innovation.









\bibliographystyle{unsrt}  
\bibliography{references,extra_ref}

\end{document}